\documentclass[fleqn,usenatbib,useAMS]{mnras}
\usepackage{graphicx}	
\usepackage{amsmath}	
\usepackage{amssymb}	
\usepackage[varg]{txfonts} 
\usepackage{caption}

\usepackage[T1]{fontenc}
\usepackage{ae,aecompl}

\DeclareMathOperator{\sech}{sech}

\def\MDM{\ifmmode{\>M_{\textnormal{\sc dm}}}\else{$$M_{\textnormal{\sc dm}}}\fi}

\def\XH{\ifmmode{\>X_{\textnormal{\sc h}}} \else{$X_{\textnormal{\sc h}}$}\fi}
\def\nH{\ifmmode{\>n_{\textnormal{\sc h}}} \else{$n_{\textnormal{\sc h}}$}\fi}

\def\maspyr{\ifmmode{\>\textnormal{mas~yr}^{-1}}\else{mas~yr$^{-1}$}\fi}

\def\mG{\ifmmode{\>\mu\mathrm{G}}\else{$\mu$G}\fi}
\def\erg{\ifmmode{\> {\rm erg}}\else{erg}\fi}
\def\keV{\ifmmode{\> {\rm keV}}\else{keV}\fi}

\def\deg{\ifmmode{\>^{\circ}}\else{$^{\circ}$}\fi}
\def\onedeg{\ifmmode{\>1^{\circ}}\else{$1^{\circ}$}\fi}

\def\xvir{\ifmmode{\>\!x_{vir}}\else{$x_{vir}$}\fi}
\def\Mvir{\ifmmode{\>\!M_{vir} }\else{$M_{vir} $}\fi}
\def\rvir{\ifmmode{\>\!r_{vir}}\else{$r_{vir}$}\fi}
\def\vvir{\ifmmode{\>\!v_{vir}}\else{$v_{vir}$}\fi}
\def\Vvir{\ifmmode{\>\!V_{vir} }\else{$V_{vir} $}\fi}

\def\tratio{\ifmmode{\>\tau}\else{$\tau$}\fi}

\def\rms{\ifmmode{\>r_{\textnormal{\sc ms}}}\else{$r_{\textnormal{\sc ms}}$}\fi}

\def\Mpc{\ifmmode{\>\!{\rm Mpc}} \else{Mpc}\fi}
\def\kpc{\ifmmode{\>\!{\rm kpc}} \else{kpc}\fi}
\def\pc{\ifmmode{\>\!{\rm pc}} \else{pc}\fi}

\def\Gyr{\ifmmode{\>\!{\rm Gyr}} \else{Gyr}\fi}
\def\Myr{\ifmmode{\>\!{\rm Myr}} \else{Myr}\fi}
\def\yr{\ifmmode{\>\!{\rm yr}} \else{yr}\fi}
\def\pyr{\ifmmode{\>\!{\rm yr}^{-1}}\else{yr $^{-1}$} \fi}
\def\s{\ifmmode{\>\!{\rm s}}\else{s}\fi}
\def\ps{\ifmmode{\>\!{\rm s}^{-1}}\else{s$^{-1}$}\fi}
\def\Hz{\ifmmode{\>\!{\rm Hz}}\else{Hz}\fi}

\def\kms{\ifmmode{\>\!{\rm km\,s}^{-1}}\else{km~s$^{-1}$}\fi}

\def\K{\ifmmode{\>\!{\rm K}}\else{K}\fi}

\def\sr{\ifmmode{\>\!{\rm sr}}\else{sr}\fi}
\def\psr{\ifmmode{\>\!{\rm sr}^{-1}}\else{sr$^{-1}$}\fi}
\def\arcs{\ifmmode{\>\!{\rm arcsec}}\else{arcsec}\fi}
\def\parcs{\ifmmode{\>\!{\rm arcsec}^{-1}}\else{arcsec${-1}$}\fi}
\def\parcss{\ifmmode{\>\!{\rm arcsec}^{-2}}\else{arcsec${-2}$}\fi}

\def\cm{\ifmmode{\>\!{\rm cm}}\else{cm}\fi}
\def\cc{\ifmmode{\>\!{\rm cm}^{3}}\else{cm$^{3}$}\fi}
\def\sqc{\ifmmode{\>\!{\rm cm}^{2}}\else{cm$^{2}$}\fi}
\def\pcc{\ifmmode{\>\!{\rm cm}^{-3}}\else{cm$^{-3}$}\fi}
\def\psc{\ifmmode{\>\!{\rm cm}^{-2}}\else{cm$^{-2}$}\fi}

\def\g{\ifmmode{\>\!{\rm g}}\else{g}\fi}
\def\Msun{\ifmmode{\>\!{\rm M}_{\odot}}\else{M$_{\odot}$}\fi}
\def\hMsun{\ifmmode{\> h^{-1}{\rm M}_{\odot}}\else{$h^{-1}$M$_{\odot}$}\fi}

\def\Zsun{\ifmmode{\>\!{\rm Z}_{\odot}}\else{Z$_{\odot}$}\fi}

\def\Lsun{\ifmmode{\>\!{\rm L}_{\odot}}\else{L$_{\odot}$}\fi}

\def\rayl{\ifmmode{\>\!{\rm R}}\else{R}\fi}
\def\mR{\ifmmode{\>\!{\rm mR}}\else{mR}\fi}

\renewcommand{\ion}[2]{\hbox{#1\,{\sc #2}}}

\def\lya{\ifmmode{\>\!{\rm Ly}\alpha}\else{Ly$\alpha$}\fi}

\def\Ha{\ifmmode{\>\!{\rm H}\alpha}\else{H$\alpha$}\fi}
\def\Hb{\ifmmode{\>\!{\rm H}\beta}\else{H$\beta$}\fi}

\def\HI{\ifmmode{\> \textnormal{\ion{H}{i}}} \else{\ion{H}{i}}\fi}
\def\HII{\ifmmode{\> \textnormal{\ion{H}{ii}}} \else{\ion{H}{ii}}\fi}
\def\CIV{\ifmmode{\> \textnormal{\ion{C}{iv}}} \else{\ion{C}{iv}}\fi}
\def\SiIV{\ifmmode{\> \textnormal{\ion{S}{iv}}} \else{\ion{Si}{iv}}\fi}

\def\NH{\ifmmode{\> {\rm N}_{\rm H}} \else{N$_{\rm H}$}\fi}
\def\Ng{\ifmmode{\> {\rm N}_{\rm gas}} \else{N$_{\rm gas}$}\fi}
\def\NHI{\ifmmode{\> {\rm N}_{\HI}} \else{N$_{\HI}$}\fi}
\def\MHI{\ifmmode{\> {\rm M}_{ \HI}} \else{M$_{\HI}$}\fi}

\def\mua{\ifmmode{\>\mu_{ \textnormal{\Ha}}}\else{$\mu_{ \textnormal{\Ha}}$}\fi}
\def\alphabha{\ifmmode{\>\alpha_{B}^{(\textnormal{\Ha})}}\else{$\alpha_{B}^{(\textnormal{\Ha})}$}\fi}

\newcommand{\ramses}{{\sc ramses}}
\newcommand{\agama}{{\sc agama}}

\newcommand{\gaia}{{\em Gaia}}

{\newif\ifnotend
\notendtrue
\def\veclist{ABCDEFGHIJKLMNOPQRSTUVWXYZabcdefghijklmnopqrstuvwxyz.}
\def\top#1#2.{#1}
\def\tail#1#2.{#2.}
\loop\expandafter\xdef\csname v\expandafter\top\veclist\endcsname%
{{\noexpand\bf\expandafter\top\veclist}}
\edef\veclist{\expandafter\tail\veclist}
\if\veclist.\notendfalse\fi\ifnotend\repeat}
\def\Gyr{\,\mathrm{Gyr}}
\def\Myr{\,\mathrm{Myr}}
\def\kpc{\,\mathrm{kpc}}

\def\d{{\rm d}}
\renewcommand{\[}{\begin{equation}}
\renewcommand{\]}{\end{equation}}

\newcommand{\Gaia}{\textit{Gaia}}

\def\HI{H${\scriptstyle\rm I}$}
\def\Msun{\ifmmode{\>\!{\rm M}_{\odot}}\else{M$_{\odot}$}\fi}
\def\Zsun{\ifmmode{\>\!{\rm Z}_{\odot}}\else{Z$_{\odot}$}\fi}
\def\XH{\ifmmode{\>X_{\textnormal{\sc h}}}\else{$X_{\textnormal{\sc h}}$}\fi}

\def\Usol {\ifmmode{U_\odot} \else {$U_\odot$}\fi} 
\def\Vsol {\ifmmode{V_\odot} \else {$V_\odot$}\fi} 
\def\Wsol {\ifmmode{W_\odot} \else {$W_\odot$}\fi} 
\def\Vsolar {\ifmmode{{\bf v}_\odot} \else {${\bf v}_\odot$}\fi} 
\def\vsolar {\ifmmode{v_\odot} \else {$v_\odot$}\fi} 
\def\VLSR {\ifmmode{{\bf \Theta}_{\rm 0}} \else {${\bf \Theta_{\rm 0}}$}\fi} 
\def\Vlsr {\ifmmode{{\bf v_{\rm LSR}}} \else {${\bf v_{\rm LSR}}$}\fi} 
\def\Rsolar{\ifmmode{R_0}\else {$R_0$}\fi}
\def\zsolar{\ifmmode{z_0}\else {$z_0$}\fi}
\def\phisolar{\ifmmode{\phi_0}\else {$\phi_0$}\fi}
\def\Omegasolar{\ifmmode{{\bf \Omega}_\odot}\else {${\bf \Omega}_\odot$}\fi}

\def\Vzmax{\ifmmode{\vert V_{z_{\rm max}}\vert}\else {$\vert V_{z_{\rm max}}\vert$}\fi}
\def\zmax{\ifmmode{\vert z_{\rm max}\vert}\else {$\vert z_{\rm max}\vert$}\fi}
\def\Az{\ifmmode{{\cal A}_z}\else {${\cal A}_z$}\fi}

\def\GBP {\ifmmode{G_{\rm BP}} \else {$G_{\rm BP}$} \fi}
\def\GRP {\ifmmode{G_{\rm RP}} \else {$G_{\rm RP}$} \fi}

\begin{document}
\label{firstpage}
\pagerange{\pageref{firstpage}--\pageref{lastpage}}

\title[The Galaxy's impulsive response]{Galactic seismology: the evolving ``phase spiral" after the Sagittarius dwarf impact}

\author[Bland-Hawthorn \& Tepper-Garc\'ia]{%
Joss Bland-Hawthorn$^{1,2}$\thanks{Contact e-mail: \href{mailto:jbh@physics.usyd.edu.au}{jbh@physics.usyd.edu.au}} and Thor~Tepper-Garc\'ia$^{1,2,3}$ \\
$^{1}$Sydney Institute for Astronomy, School of Physics, A28, The University of Sydney, NSW 2006, Australia\\
$^{2}$Center of Excellence for All Sky Astrophysics in Three Dimensions (ASTRO-3D), Australia\\
$^{3}$Centre for Integrated Sustainability Analysis, School of Physics, The University of Sydney, NSW 2006, Australia
}

\pubyear{2020}

\maketitle

\begin{abstract}
{\bf
In 2018, the ESA \Gaia\ satellite discovered a remarkable spiral pattern (``phase spiral") in the $z-V_z$ phase plane throughout the solar neighbourhood, where $z$ and $V_z$ are the displacement and velocity of a star perpendicular to the Galactic disc. In response to Binney \& Sch\"onrich's analytic model of a disc-crossing satellite to explain the \Gaia\ data,
we carry out a high-resolution, N-body simulation (N$\:\approx 10^8$ particles) of an impulsive mass ($2\times 10^{10}$ \Msun) that interacts with a cold stellar disc at a single transit point. The disc response is complex since the impulse triggers a superposition of two distinct bisymmetric ($m=2$) modes $-$ a density wave and a corrugated bending wave $-$ that wrap up at different rates. Stars in the {\it faster} density wave wrap up with time $T$ according to $\phi_D(R,T)=(\Omega_D(R) + \Omega_{\rm o})\:T$ where $\phi_D$ describes the spiral pattern and $\Omega_D =\Omega(R) -\kappa(R)/2$, where $\kappa$ is the epicyclic frequency. While the pattern speed $\Omega_{\rm o}$ is small, it is non-zero. The {\it slower} bending wave wraps up according to $\Omega_B\approx\Omega_D/2$ producing a corrugated wave. The bunching effect of the density wave triggers the phase spiral as it rolls up and down on the bending wave (``rollercoaster'' model). The phase spiral emerges slowly about $\Delta T \approx 400$ Myr after impact.
It appears to be a long-lived, disc-wide phenomenon that continues to evolve over most of the 2~Gyr simulation. Thus, given Sagittarius' (Sgr) low total mass today ($M_{\rm tot}\sim 3\times 10^8$ \Msun\ within 10 kpc diameter), we believe the phase spiral was excited by the disc-crossing dwarf some $1-2$ Gyr {\it before} the recent transit. For this to be true, Sgr must be losing mass at 0.5-1 dex per orbit loop.
}
\end{abstract}
\begin{keywords}
Surveys -- 
the Galaxy --
stars: dynamics --
stars: kinematics --
methods: N-body simulations --
methods: analytic
\end{keywords}

\section{Introduction} \label{s:intro}

In April 2018, the ESA \Gaia\ astrometric mission \citep{perryman2001,prusti2016,soubiran2018} presented its second data release (DR2). These observations have revolutionized both stellar and Galactic studies \citep[DR2:][]{Brown2018}. After only two years of
observations, the remarkable precision of \Gaia's measured stellar parameters has led to new discoveries and new fields of study. These data are being complemented {\it inter alia} by large spectroscopic surveys -- 
RAVE \citep{Steinmetz2006}, APOGEE \citep{Majewski2017}, 
\Gaia-ESO \citep{Gilmore2012}, LAMOST \citep{Deng2012} and GALAH \citep{deSilva2015}. In all respects, this is the golden
age of Galactic archaeology \citep{freeman2002}.

{\bf One of the most important discoveries to emerge from the ESA \Gaia\ astrometric survey is an unexpected phase-space signal in the local stellar disc \citep{antoja2018}. 
In Galactic cylindrical coordinates ($R,\phi,z$), individual stars have velocities ($V_R$, $V_\phi$, $V_z$) and oscillation frequencies
($\Omega_R,\Omega_\phi,\Omega_z$) $=$ ($\kappa,\Omega,\nu$).
}
In a volume element
defined by ($\Delta R, R\,\Delta \phi, \Delta z$) = ($\pm 0.1$, $\pm 0.1$, $\pm 1$) kpc$^3$
centred on the Sun, the \gaia\ team detect a coherent spiral
pattern\footnote{This feature has been variously referred to as the ``snail" \citep{antoja2018}, the ``snail shell" \citep{Li2020,Li2021}, the ``Gaia spiral" \citep{Darling2019a}, the phase-plane spiral \citep[][]{Binney2018} and the phase-space spiral \citep[][]{Michtchenko2019}. Consistent with the traditional use of ``phase mixing'' rather than ``phase-space mixing,'' we adopt the term ``phase spiral" \citep[e.g.][]{Khanna2019,Hunt2019,Mackereth2019,Wang2019}.
The advantage of the more compact language becomes apparent when used as an adjective, e.g. phase-spiral evolution, phase-spiral dynamics, etc.} in the phase plane defined by $z$ and $V_z$. 
The phase spiral is most evident when each point of the $z-V_z$ phase plane is represented by either $\langle V_R\rangle$ or $\langle V_\phi\rangle$, averaged over the local volume. {\bf (Thereafter, \citealt{Khanna2019} showed the phase spiral emerges more clearly when encoded by $\langle L_z\rangle$, which measures stellar angular momentum about the Galaxy's spin axis.)}
This phenomenon is indicative of a system that is settling from a mildly disturbed state to a stationary configuration through the process of phase mixing \citep{LyndenBell1967}.

Since its recent discovery,
up to a dozen independent research papers
seek to explain or shed light on the phase spiral phenomenon.
Explanations to date invoke three different mechanisms: (i) partial coherence imposed by the recent dissolution of several star clusters \citep[][]{Candlish2014,Michtchenko2019}; (ii) vertical oscillations excited by the strong buckling of a stellar bar \citep[][]{Khoperskov2019}; and (iii) vertical and in-plane 
oscillations induced by a massive disc-crossing perturber
\citep{Binney2018,Darling2019a}.
The first two mechanisms are of internal origin; the third external mechanism has received the majority of attention
to date, not least because the Sgr dwarf galaxy is observed to be undergoing disruption as it crosses the disc. This is the focus
of our attention because we consider it to be the most
likely interpretation for the phase-spiral phenomenon
\citep{Laporte2019,BlandHawthorn2019}.
The effect is now observed over too large a surface area (see below) to be a coherent phenomenon arising from dissolving star clusters. Moreover, the extremely tight alignment of the plane of the stellar bar with the inner disc plane \citep{blandhawthorn2016} imposes strong constraints on the idea of a buckling bar. 

Three independent observing campaigns confirm the phase-space pattern and study it in more detail. With the aid of the GALAH survey, \citet[][hereafter B19]{BlandHawthorn2019} dissected the phase spiral as
a function of orbit actions, stellar ages and stellar abundances. They were able to detect the pattern over a larger radial range ($7 < R < 9$ kpc) and azimuthal angle ($\delta\phi \lesssim \pm 15^\circ$). The LAMOST survey provides some evidence
for a changing phase-spiral pattern over the radial range $6 < R < 12$ kpc, as predicted in action-based, analytic models (e.g. B19, their Fig. 20) and observed in the highest contrast numerical simulation to date \cite[][hereafter L19]{Laporte2019}.
Both \citet[][their Fig. 4]{Li2020} and B19 present evidence that the phase spiral is {\bf more clearly defined} in the cold orbits and less clear in the hotter, $\alpha$-enhanced disc, further underscored by the 2-component disc model in B19.

In these studies, the spiral pattern in phase space is found to be somewhat patchy and uneven. This is due in part to the well-documented kinematic \citep[][]{Widrow2012,Williams2013,Carlin2013} and density asymmetries \citep[][]{Yanny2013,Slater2014,Xu2015}
when comparing the upper and lower sides of the Galactic plane. However, when the $z-V_z$ plane is encoded as the angular momentum about the $z$-axis, $\langle L_z\rangle$ (action $\langle J_\phi\rangle$), the phase-spiral pattern is smooth and varies systematically as a function of disc location \citep{Khanna2019}. This suggests that the emergence of the phase spiral is somehow related to the angular momentum properties of the disc, in accord with the BS18 model.

Galactic seismology\footnote{For a brief history of this emerging field, see Appendix~\ref{s:history}. The terminology is used interchangeably with galactoseismology \citep{Widrow2012}.},
a subset of Galactic archaeology, is useful for exploring past interactions and structural properties of the Galaxy. Specifically, we seek a better understanding of how a differentially-rotating stellar disc responds to a powerful impulse and how this sets up the phase-spiral phenomenon. 
{\bf 
A key question is what are the correct properties to ascribe to Sgr $-$ its total mass, infall velocity, gas fraction and impact radius $-$ {\it at a specific epoch in the past} \cite[e.g.][]{Vasiliev2020,Vasiliev2021}. In our view, new observations indicate
that Sgr's original mass, presumably prior to infall, is likely to have been comparable to the LMC at the present epoch. In the core of the dwarf, the 
high metallicity tail in [Fe/H] 
\citep[][their Fig. 14]{Nidever2020} extends well beyond the SMC metal distribution and is comparable to, or exceeds that, of the LMC. The metallicity spread is indicative of the high-mass limit for dwarf galaxies \citep{Lee2008,Hayes2020}.
Its star formation history \citep{RuizLara2020} indicates that Sgr lost its gas on successive disc crossings, as modelled for the first time by \cite{TepperGarcia2018}. Sgr's present total mass is about $2-4\times 10^8$ \Msun\ within a radius of 5 kpc from the dwarf's centre \citep{Law2005,Vasiliev2020}.
Thus, Sgr's original total mass and baryon content (Sec.~\ref{s:disc}) must have been $1-2$ orders of magnitude higher at some time in the distant past
\cite[][their Fig. 4]{Jiang2000}. 

For consistency with most earlier work, especially \cite{Binney2018}, we adopt an initial perturber mass of $2\times 10^{10}$ \Msun\ to drive the impulse \citep[see also][]{Gomez2013}. Earlier work has already shown that this is the low-mass threshold needed to drive the necessary response (L19, B19); models using a lower perturber mass can account for Sgr's tidal tails \citep[e.g.][]{Law2005} but produce a very weak vertical disc response. We do not view the apparent inconsistency with observations as a problem. If the phase spiral phenomenon is long-lived, a disc crossing prior to the current event may in fact be responsible. But this raises new questions about the stripping rate experienced by Sgr as a function of time, issues we address in Sec.~\ref{s:disc}.
}

In what follows, we start by reviewing the \citet[][hereafter BS18]{Binney2018} model (Sec.~\ref{s:BS18}),
which provides the impetus
for our new work, before discussing the overall strategy (Sec.~\ref{s:motiv}).
The main goal is to design and carry out a numerical simulation that bridges between the simplicity of an analytic model (BS18)
and the complexity of a realistic  numerical simulation (e.g. L19).
In Sec.~\ref{s:nbody}, we discuss the adopted parameters of the Galactic model and the intruder. 
The Galactic model is set up carefully with specific initial conditions that guarantee long-term stability in an N-body simulation. In Sec.~\ref{s:analy}, 
with reference to web links to our simulations, we
describe our analysis and the new results. In the final discussion (Sec.~\ref{s:disc}), we show that the new simulations are a major improvement over earlier work (L19, B19), although many questions remain.

\section{Binney \& Sch\"onrich (2018) model} \label{s:BS18}

BS18 develop a purely analytic (toy) model to investigate the response of the Galactic stellar disc to the perturbation induced by an external massive object. They employ action-based modelling defined by distribution functions $f(\mathbf J)$ that are analytic functions of
the action integrals ${\mathbf J} =(J_R, J_\phi, J_z)$. With recourse to the Action-based GAlaxy Model Architecture (\agama) package \citep{Vasiliev2019}, they produce a highly self-consistent, multi-component model of the Galaxy. Their vertical density profile through the solar neighbourhood is an excellent match to the observed two-component disc profile \citep{Gilmore1983}. This is important if the model is to match the vertical actions ($J_z$) and frequencies ($\Omega_z\equiv\nu$) given the anharmonic nature of the vertical gravitational potential. They choose phase-space coordinates for a million stars in the solar neighbourhood and backwards integrate them
to a time when the perturber first appears. The intruder
is moving with a downward velocity of $\vec{v}_P \approx 300$ \kms, has a mass of $M_P = 2\times 10^{10}$ \Msun\ and impacts the disc at $R_P=18$ kpc along the Sun's radius vector. They are then able to compute the acceleration imposed by the intruder on the particles relative to the unperturbed acceleration. 

BS18 observe that the largest off-planar perturbation is expected at $\phi-\phi_o \sim \pm \pi/2$ because these stars are either moving towards ($\phi-\phi_o < 0$) or away ($\phi-\phi_o > 0$) from the perturber as it transits the disc. (Here, $\phi_o=0$ because the intruder transits along the Sun's radius vector.) Thus, stars approaching $\phi=\phi_o$ feel a net upward pull, and stars receding from $\phi=\phi_o$ feel a net downward pull. This action sets up a strong low-order bending mode across the disc.
As the perturber transits the disc, stars near the crossing point experience a strong in-plane tug towards the perturber's trajectory, which binds the in-plane and vertical actions.

\begin{figure}
\centering
\includegraphics[width=0.5\textwidth]{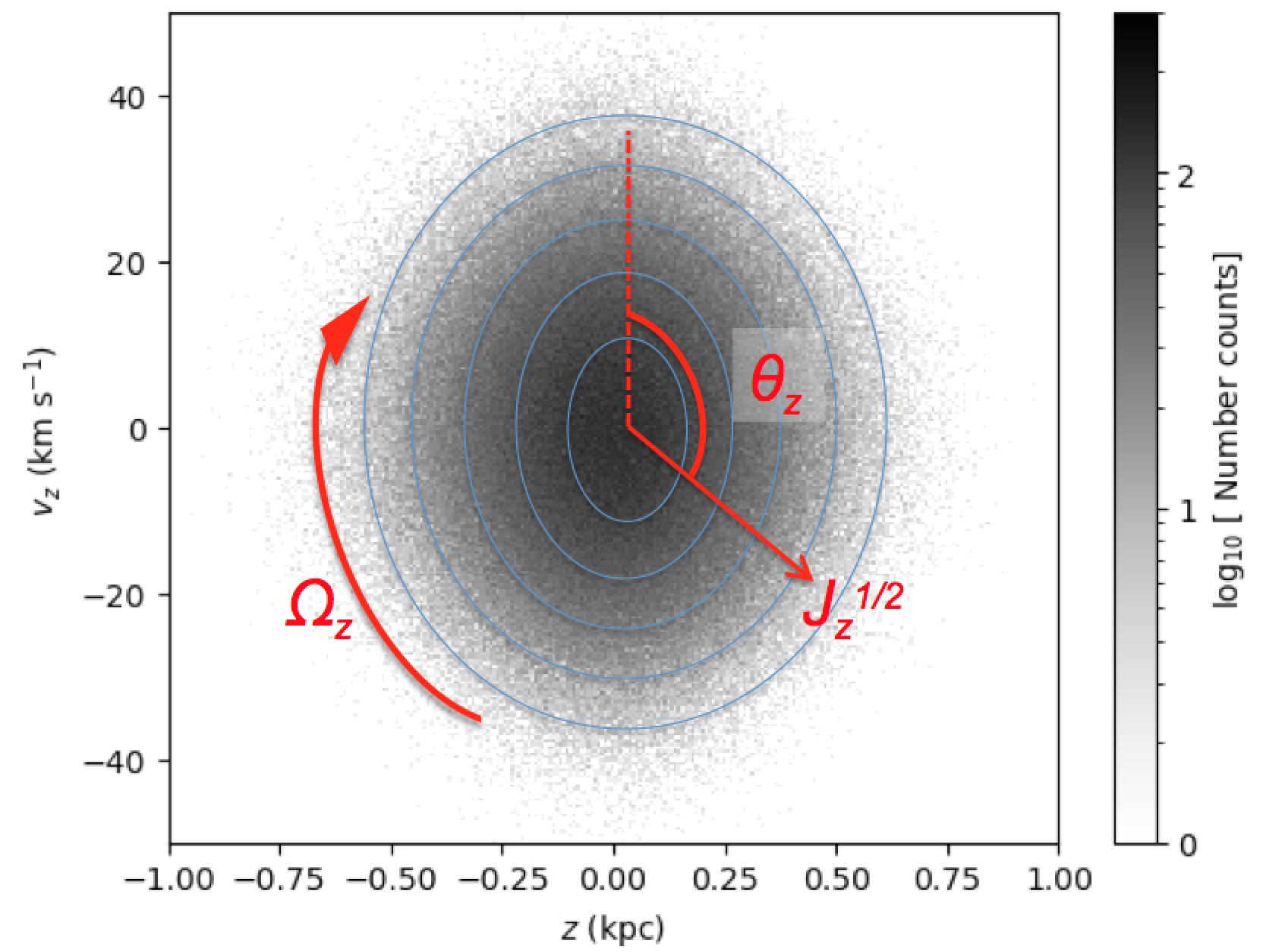} \caption{The natural coordinate system in the $z-V_z$ 
plane: the conjugate angle $\theta_z$ is the angle from the vertical; the radial distance of a star from the centre of this plane is proportional to the action $\sqrt{J_z}$. These are overlaid onto the projected stellar density in the solar neighbourhood where the simulated gaussian distribution (with random phases) arises from the vertical dispersion $\sigma_z$ profile and the scale height of the disc (Sec.~\ref{s:nbody}).
The blue ellipses are surfaces of section for orbits with a fixed vertical action $J_z$. The vertical disc frequency $\Omega_z$ of the orbit determines the rate at which a star moves around the centre $(z,V_z)=(0,0)$.
}
\label{f:zVz}
\end{figure}

\begin{figure}
\centering
\includegraphics[width=0.5\textwidth]{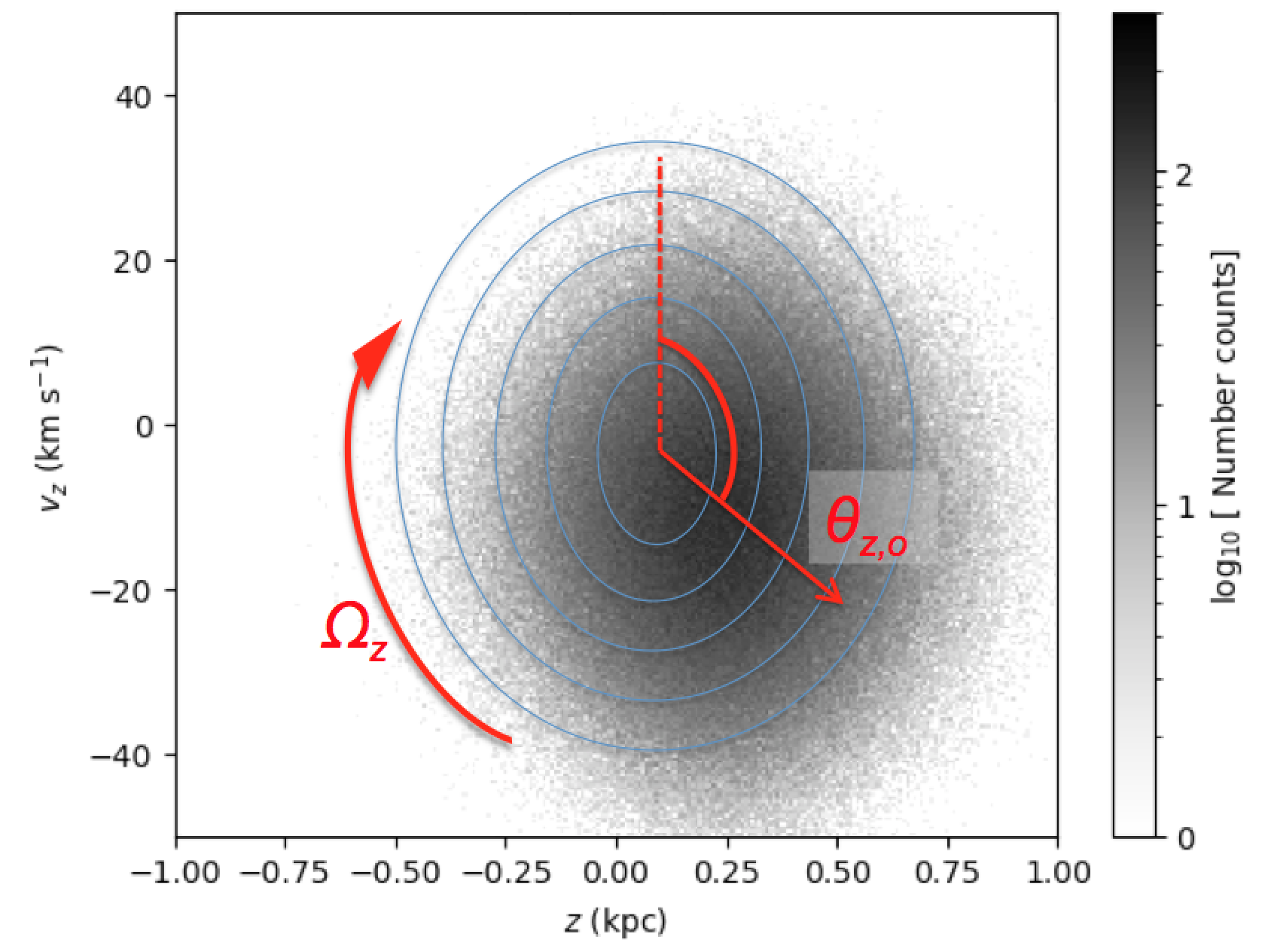} \caption{In the BS18 model, the impact causes the local stellar distribution to ``slosh'' in the $z-V_z$ plane with respect to the underlying large-scale potential.
In contrast to Fig. 1, the disturbance has imposed some phase coherence on the stellar population. Each star now has a different vertical frequency $\Omega_z$ compared to its position in the unperturbed distribution and starts to circulate out of equilibrium (phase mix) in the $z-V_z$ plane. 
}
\label{f:zVz-slosh}
\end{figure}

\begin{figure}
\centering
\includegraphics[width=0.45\textwidth]{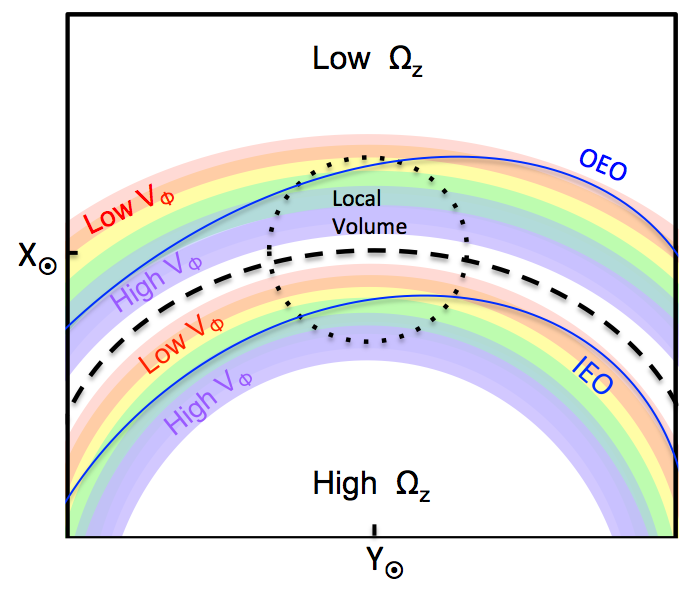} \caption{Stars from the inner and outer disc are able to reach the solar neighbourhood along elliptic orbits. In a righthanded Galactocentric coordinate frame, the Sun is at $(X_\odot,Y_\odot)=(-8.2,0)$ kpc, as indicated. 
We show an inner elliptic orbit (IEO) and an outer elliptic orbit (OEO); these precess about the Galactic Centre but remain confined to an annulus about the guiding radius ($R_g$) at their midpoint.
In the BS18 model, the tangential velocity $V_\phi$ of a star in the Local Volume is intimately associated with the vertical frequency $\Omega_z$ of the star's orbit. 
Stars arriving from the inner disc typically have lower $V_\phi$ but higher $\Omega_z$ due to the higher disk surface density;
the converse is true for stars arriving from the outer disc. Stars from the inner disc circulate faster in the $z-V_z$ plane (Fig.~\ref{f:zVz}) compared to stars from the outer disc. (The chosen colour scheme is consistent with BS18.)
}
\label{f:BS18}
\end{figure}

With the aid of Figs.~\ref{f:zVz}-\ref{f:BS18}, we illustrate a key insight from the BS18 model. The coordinate system of the $z-V_z$ plane is shown in Fig. ~\ref{f:zVz} superposed on the surfaces of section for different stellar orbits confined to tori about the Galactic Centre. In Fig.~\ref{f:zVz-slosh}, sloshing in the $z-V_z$ plane induced by the impact leads to a stellar overdensity that is offset with respect 
to the original stratification at a mean angle of $\theta_z \approx \theta_{z,o}$ (Fig. ~\ref{f:zVz-slosh}). Whereas before stars had constant values of $J_z$ that were independent of the conjugate angle
$\theta_z$, the entire distribution now depends on $\theta_z$ and the
assigned $J_z$ associated with the new ellipse.
Since for a disc potential, $\Omega_z$ declines with $\sqrt{J_z}$ \citep[e.g.][BS18]{Candlish2014}, a  phase-spiral structure starts to develop. Crucially, BS18 note that the phase spiral is not apparent in stellar overdensity, although L19 detect a weak signal, in agreement with the original study \citep{antoja2018}. {\bf Therefore, a different} mechanism is needed for the
pattern to emerge so clearly in $\langle V_\phi\rangle$ and $\langle V_R\rangle$,
{\bf which we now discuss.}

In Fig.~\ref{f:BS18}, we illustrate how elliptic orbits from the inner
and outer disc are able to enter the local volume. The vertical oscillation
frequency $\Omega_z$ declines smoothly with increasing radius because
of the disc's exponential decline in the local surface density. Stars from the inner disc reach us at their aphelia, and these are characterised by lower $\langle V_\phi\rangle$ and higher $\langle\Omega_z\rangle$; the converse is true for stars coming in from the outer disc.

{\bf 
In Fig.~\ref{f:zVz-slosh}, the dislodged stars now have new positions in the $z-V_z$ plane. At every point in that plane, stars have a spread of orbit properties where their clockwise movement in the plane is determined by their current $\Omega_z$ value. The dislodged stars trace ellipses in the $z-V_z$ plane with a conjugate angle that grows with time $t$ as $\theta_z = \Omega_z t + \theta_{z,o}$.
But $\Omega_z$ is tied to $\langle V_\phi\rangle$ (and $\langle V_R\rangle$) such that stars with lower $\langle V_\phi\rangle$ move ahead of stars with higher $\langle V_\phi\rangle$. Thus, the phase spiral is defined by a change in $\langle V_\phi\rangle$ (and $\langle V_R\rangle$)
as we move clockwise around each ellipse in the $z-V_z$ plane. The change in projected $\langle V_\phi\rangle$ (and $\langle V_R\rangle$) becomes decoupled from the projected density along the ellipses in Fig.~\ref{f:zVz-slosh}. This is why the phase spiral (with its intrinsically higher dynamical range) emerges so clearly when encoded kinematically, and is visible even when the projected density of stars is comparatively low (see Sec.~\ref{s:PSevol}).
}

\section{Motivation \& Strategy} \label{s:motiv}

The motivation for our new work can be stated as follows. We seek to understand the phase-spiral phenomenon with the simplest possible model that is true to the BS18 toy model while incorporating the advantages of the N-body approach. First, our work
demands a more rigorous treatment for setting up the initial conditions. We require the coldest possible synthetic disc that comes into equilibrium at the start of the simulation and does not develop significant substructure or change its properties over the course of many disc rotations. \cite{Darling2019a} demonstrate that the emergence of the phase spiral is a competition between phase mixing and the disc's self-gravity. Consistent with our approach, a lower self-gravity leads to a stronger signal (higher contrast with respect to the background population) although a higher self-gravity appears to sustain the signature for longer.

Secondly, we require a better treatment of the impulse approximation by an N-body solver. In particular, the heavy point mass must be `live' on its first and only approach, and during the transit, and thereafter it must not continue to disturb the disc.
In our earlier work (B19), the disc response can be more erratic in the post-transit phase due to the sustained force applied by the intruder.
While there is an in-built asymmetry to our approach, this is true of all Sgr-based orbit families because of dynamical friction and the perturber's mass loss along its orbit \citep{Law2005,Nichols2014}.
The underlying mechanisms involved in the disc response to an extended,
disrupting perturber are obfuscated by many overlapping processes 
(e.g. L19, B19). 

In contrast,
the BS18 toy model uses a theoretical description of the Galaxy that is analysed in terms of distribution functions defined in action space, as discussed in Sec.~\ref{s:BS18}.
The latter model, while computationally efficient, lacks the realism of 
the actual system, most notably the complex recoil of both Sgr and the Milky Way to their mutual attraction and, importantly, the self-gravity and `elasticity' of the dominant stellar disc. 
How the vertical and in-plane motions emerge and work together during the interaction are unclear \citep[][]{dOnghia2016,Darling2019a,Poggio2021}. But by using a model that bridges the divide between analytic and N-body models, we seek to shed more light on this process.

If we are to compare the BS18 model meaningfully with an N-body simulation, the basic parameters of their set-up must be matched closely (Sec.~\ref{s:model}). In our model, a point mass is used in place of an extended perturber to maintain the comparison and is placed on a hyperbolic orbit to minimize ongoing interactions with the disc.
\cite{Aguilar1985} note that in order to recreate the impulse approximation within an N-body simulation, the interval during which the mutual gravitational forces are significant is short compared to the crossing time within each system \citep[see also][\S8.2]{Binney2008}. 
The ``impulsiveness'' is further ensured (see Sec.~\ref{s:impulse}) by a perturber mass that is constant until the point of disc transit, and then declines exponentially in time such that the mass is negligible at the second disc crossing (Sec.~\ref{s:model}).
In the next section, we explore a new suite of models to simulate a clean impulse imparted by a disc-crossing perturber. 

In order to achieve the fidelity of the phase spiral in the BS18 models, these authors recommend that any N-body simulation use $\sim 10^8$ disc particles, an order of magnitude larger than any matched, impulsive disc simulation to date.
BS18 observe a synthetic phase spiral over an area in the $z-V_z$ plane that is comparable to \cite{antoja2018}, i.e. $\pm 60$ \kms\ and $\pm$1 kpc, a fourfold improvement over L19 and B19. After following BS18's recommendation, we also achieve phase spiral patterns over the smaller region.

\section{N-body simulations} \label{s:nbody}

\subsection{The model} \label{s:model}

\begin{table*}
\begin{center}
\caption{Galaxy model parameters: 
In columns 1 and 2, 
we show the Galactic component and the intended functional forms; these are only approximate because they share the same gravitational potential. In column 3, the function variable is shown where ($R,z$) are cylindrical coordinates, and $r$ is the spherical radius. The total mass, scale length and cut-off radius are indicated in columns 4, 5, 6 respectively. Column 7 is the number of collisionless particles used in the simulation.
}
\label{t:comp}
\begin{tabular}{lllcccc}
Component & Function & Variable & Total mass & Radial scale length & Cut-off radius & Particle count \\
 &  & & $M_{\rm tot}$ ($10^{10}$ \Msun) & $r_s$ (kpc) & $r_c$ (kpc) & $N$ ($10^6$) \\
\hline
DM halo & NFW & $r$ & 140 & 15 & 300 & 20 \\
Stellar bulge & Hernquist & $r$ & 1.5 & 0.6 & 2.0 & 4.5 \\
Stellar disc & Exp, $\sech^2$ & $R, z$ & 3.4 & 3.0 & 40 & 50\\
\hline
\end{tabular}
\end{center}
\begin{list}{}{}
\item Notes: The NFW and Hernquist functions are defined elsewhere \citep[][]{NFW1997,her90a}. The scale height of the stellar disc is $z_t \approx 250$ pc; the local instability parameter is everywhere $Q \gtrsim 1.3$ \citep{too64a}.
\end{list}
\end{table*}

\begin{figure*}
\centering
\includegraphics[width=0.45\textwidth]{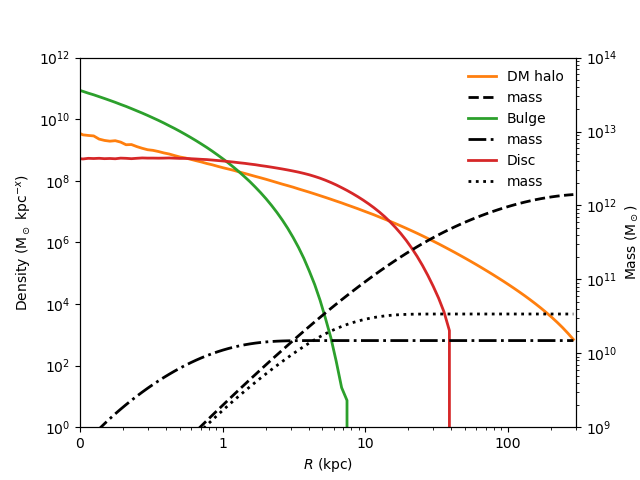}
\includegraphics[width=0.45\textwidth]{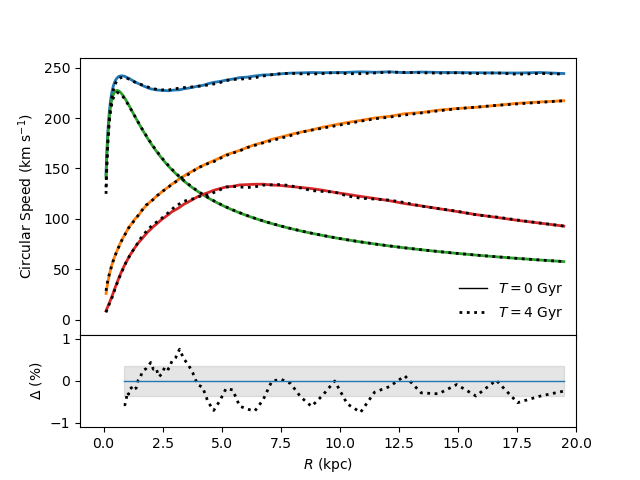}
\includegraphics[width=0.45\textwidth]{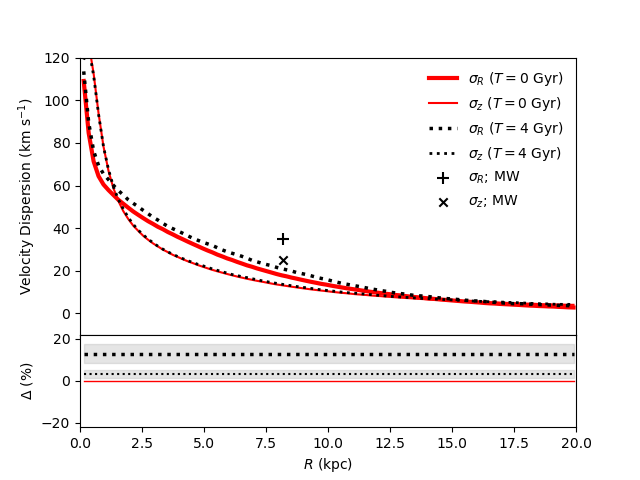}
\includegraphics[width=0.45\textwidth]{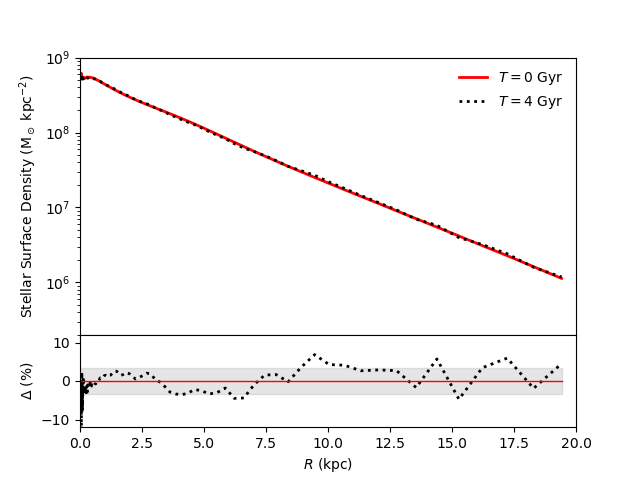}
\caption{(Top left) The density and integrated mass profiles of our three-component Galaxy model. The curves corresponding to the bulge's and disc's density distribution truncate because of the limited spatial extension of these components. The DM halo extends out to 300 kpc. Note that the density distribution of the halo and bulge are given in $\Msun~\kpc^{-3}$ (volume density), while the disc's density distribution is given in $\Msun~\kpc^{-2}$ (surface density). (Top right) The flat rotation curve (blue) of the synthetic Galaxy taken from the N-body simulation showing an acceptable resemblance to the Milky Way \citep{blandhawthorn2016}. The contributions to the rotation are also shown: bulge (green dot-dashed line), disc (red dotted line), dark halo (orange dashed line). The sub-panel shows the relative change (in percent) in the total rotation curve between these time frames. The gray-shaded area indicates the rms value around the mean. (Bottom left) The radial $\sigma_R$ (solid) and vertical $\sigma_z$ (dot-dashed) velocity dispersion profiles of the cold, synthetic disc extracted from the N-body simulation. In comparison to the Galaxy \citep{blandhawthorn2016}, $\sigma_R \approx 35$ \kms\ ($+$) and $\sigma_z \approx 25$ \kms\ ($\times$) at the solar circle ($R_\odot = 8.2$ kpc), as indicated. The sub-panel shows the {\em mean} relative change (in percent) in each of the velocity dispersions between these time frames (horizontal lines: $\Delta(\sigma_R) \approx 13$\%; $\Delta(\sigma_z) \approx 3$\%) and the corresponding rms value around the mean (gray-shaded area).  (Bottom right) The surface density of the disc: note that the red line is identical to the red curve in the top left panel. The sub-panel displays the relative change in the surface density between these two time frames. The gray-shaded area indicates the rms value around the mean.
}
\label{f:struct}
\end{figure*}

We conduct a series of N-body simulations using the tried and tested \ramses\ simulation package \citep[version 3.0 last described by][]{tey02a}. The Galaxy is approximated by a three-component system consisting of a dark-matter (DM) host halo, a stellar bulge, and a cold stellar disc; a complete list of parameters is given in Tab.~\ref{t:comp}. The stellar disc is sampled with $5\times10^7$ particles, the largest 
impulsive model of this kind to date. The model Galaxy has properties that are consistent with their observed counterparts \citep[q.v.][]{blandhawthorn2016}. Its total mass is $M_{\rm tot} \approx 1.45 \times 10^{12}$ \Msun\ with a distribution that yields a nearly flat rotation curve out to $R = 20$ kpc; the tangential velocity remains bounded within 240 to 250 \kms\ (Fig.~\ref{f:struct}, top right). The density and integrated mass profiles of each component are also shown (top left),
along with  the surface density profile of the disc alone (bottom right).

The LAMOST and GALAH studies reveal that the phase spiral is mostly confined to stellar populations with the coldest orbits \citep[B19,][]{Li2020}.  In order to enhance the contrast of the phase-spiral signature in the stellar component, we achieve one of the coldest, long-lived disc simulations to date (Sec.~\ref{s:model}). The 
disc is stabilised by our choice of
\citet[][]{too64a}'s stability parameter
\begin{equation}
Q_{\star} \equiv  {\sigma_R \, \kappa  \over 3.36 \, G \,
\Sigma_{\star} },
\end{equation}
where $\kappa$ and $\Sigma_{\star}$ are the epicyclic frequency and stellar
surface density,that provides local axisymmetric stability if we ensure $Q\gtrsim 1.3$ throughout the disc.
Our choice of $Q$ and disc mass fraction (Sec.~\ref{s:stab}) ensures that no bar or spiral arms form in isolation over a time frame of 4 Gyr \citep[][see Sec.~\ref{s:stab} below]{Fujii2018}.
These complicating factors, that were not considered by BS18, are
deferred to a follow-up paper.
The stellar dispersion profiles $\sigma_R$($R$) and $\sigma_z$($R$) are presented in Fig.~\ref{f:struct} (bottom left); these apply to the unperturbed Galaxy model in isolation, but they are identical to the corresponding properties of the interacting Galaxy at the start of the simulation.

In line with BS18, the Sagittarius dwarf galaxy (Sgr) is represented by a point-like object. The point mass is placed on a hyperbolic orbit that intersects the Galactic plane at $R\approx 18$ kpc, at which point Sgr is travelling at $\sim 330$ \kms. These constraints are consistent with Sgr's trefoil orbit parameters at the last crossing \citep[e.g.][]{TepperGarcia2018}. The motivation for using a hyperbolic orbit is to limit the interaction between the Galactic disc and the point mass to a single impact (impulse approximation) and then trace the disc's
evolution over a timespan of up to 2 Gyr. This extended time frame is in line with earlier work that shows the phase spiral may be long-lived \citep[][B19]{Darling2019a}.

The adopted orbital parameters given above imply that the perturber is moving on a bound orbit, and must therefore cross the disc more than once in the allotted time frame. To circumvent this problem, we impose on Sgr the following mass evolution:
\begin{equation*}  
M_{\rm P}(t) = M_{\rm P}(0) \times \left\{
        \begin{array}{ll}
            1 & \quad t < t_\circ \\
            \exp\left[{-(t - t_\circ)/\tau_s}\right] & \quad t \geq t_\circ
        \end{array}
    \right.
    \label{e:mexp}
\end{equation*}
where $M_{\rm P}(0) = 2\times10^{10}$ \Msun\ and $\tau_s = 30$ Myr, roughly a disc crossing time. The first disc crossing happens at $\sim 100$ Myr so we set $t_\circ = 150$ Myr. Thus the intruder mass remains constant until 40$-$50 Myr after the first disc transit, and declines by a factor of $\sim 10^6$ at the time of the second disc crossing ($t\approx 450$ Myr). The perturber's post-transit impact on the disc is entirely negligible. We experimented with these parameters and found that the impulse approximation is robust to our assumptions (Sec.~\ref{s:impulse}).

\subsection{Initial conditions and evolution} \label{s:ics}
 
In this work, we pay particular attention to the initial conditions (ICs) that specify our N-body models at the outset. 
A recurrent problem with N-body simulations to date, with rare exceptions \citep{Widrow2005}, is that the initially specified, multi-component system evolves to a new configuration before long-term stability is achieved; the 
simulator must then accept a model that is less than ideal and not what was specified.
This has been a longstanding problem with numerical simulations that the \agama\ code originally set out to resolve \citep{Vasiliev2019}. 

We exploit \agama\ to generate the positions and velocities for all components of our model (Table 1) in a self-consistent fashion.
This guarantees that
the ICs are virtually stationary from the outset\footnote{To encourage more extensive use, our \agama\ setup files are available upon request.}.

The ICs of the perturber are calculated using a backwards integration scheme in the two-body approximation where the Galaxy is described by an extended rigid body \citep[for details, see][]{TepperGarcia2020}. This is subject to the condition that the point mass crosses the Galactic plane after 95 Myr at a Galactocentric distance of 
$x = -18$ kpc with its velocity vector perfectly perpendicular to the Galactic plane and a magnitude of 330 \kms.  This is a righthanded Galactic coordinate system (Fig.~\ref{f:BS18}) with the Sun at ($x_\odot,y_\odot)=(-8.2,0)$ kpc, which has become the standard in recent years \citep[q.v.][]{Vasiliev2020}.
The initial position and velocity of the perturber with respect to the Galactic Centre thus obtained are $\vec{r}_{\rm P} \approx (-11.1,0,28.6)$ and $\vec{v}_{\rm P} = (-145.4,0,-220.2)$ respectively.

The compound (Galaxy $+$ point mass) N-body ICs are evolved with the \ramses\ code \citep{tey02a}, which incorporates adaptive mesh refinement (AMR). The system is placed into a cubic box spanning 600~kpc on a side. The AMR grid is maximally refined up to level 14, implying a limiting spatial resolution of $600 ~\kpc / 2^{14} \approx 36 ~\pc$. 
{\bf The refinement maps at a fixed timestep for both the isolated and interacting discs are shown at our main website; see footnote~\ref{F:movie}. Here we see that the maximum resolution is achieved everywhere within $R\approx 10$ kpc, i.e. the domain of our study.}
The total simulation time is 2 Gyr for the phase-spiral enactments.
We also explored a more expensive simulation with 2 pc resolution, but there were no perceptible benefits for our phase-spiral study, so we did not proceed further.\footnote{Our \ramses\ setup files are available upon request.}

Due to the presence of the simulation mesh, the force exerted by (and onto) any particle is necessarily softened, at least by a length equivalent to the side of cubic volume element (cell) equal to the limiting spatial resolution. We choose to soften the force exerted by the point-like perturber using a length equal to two such elements ($\epsilon_s \approx 72$ pc); for all other particles, $\epsilon_s$ is half this value. In addition, the force between the perturber and all the other particles is calculated by a direct N-body solver, rather than the particle-mesh (PM) method used otherwise.

\begin{figure*}
\centering
\includegraphics[width=0.45\textwidth]{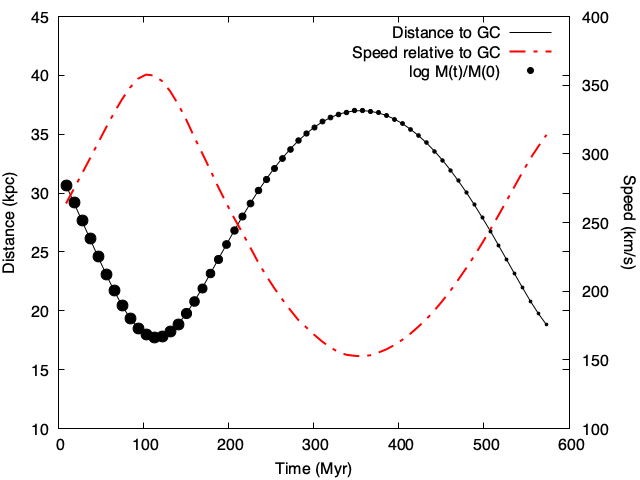}
\includegraphics[width=0.45\textwidth]{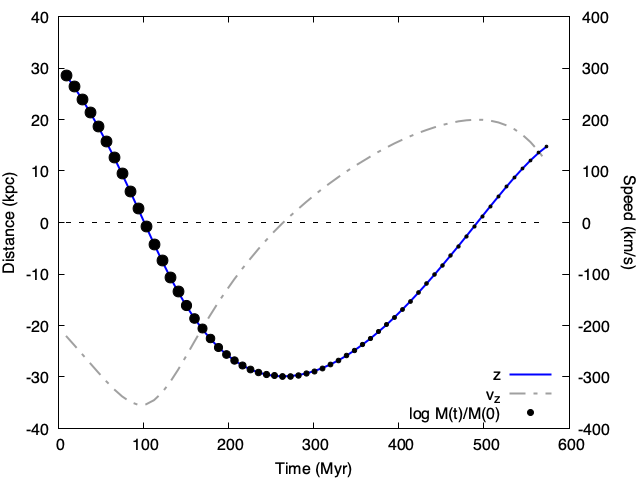}
\caption{Orbital history of a point mass perturber (Sgr) around the Galaxy in the N-body simulation. (Left) Total distance (black solid curve) and speed (red dot-dashed curve) relative to the Galactic Centre. (Right) Vertical distance $z$ (blue solid curve) and vertical speed $V_z$ (gray dot-dashed curve) relative to the Galactic plane. The disc crossings are identified by the intersection of the blue curve and $z = 0$ (dashed horizontal line) and are clearly visible at $T \approx 100$ Myr, and $T \approx 500$ Myr (bottom). The black dots on top of the black curve (top) and of the blue curve (bottom) are indicative of Sgr's mass along its orbit. The size of the symbol is directly proportional to $\log_{10}$(mass) relative to its initial value.
}
\label{f:orbit}
\end{figure*}

From its starting point above the Galactic plane, the perturber reaches the stellar disc travelling at about 350 \kms\ after 95 Myr from an initial galactocentric distance of about 40 kpc above the disc. We provide a link to the movie in the footnote.\footnote{\label{F:movie}
\url{http://www.physics.usyd.edu.au/~tepper/proj\_galseismo.html}}
It crosses the disc at $R \approx 18$ kpc, i.e. $(x,y) \approx (-18,0)$ kpc, and transits below the Galactic plane to a vertical distance of about $z\approx 35$ kpc some 250 Myr after the disc crossing, then falls back towards the Galaxy $-$ see Fig.~\ref{f:orbit}.
At $t_\circ = 150$ Myr, i.e. shortly after it crosses the Galactic plane, we artificially decrease its mass over an e-folding timescale, $\tau_s = 30$ Myr (Sec.~\ref{s:model}), such that by the time the intruder crosses the disc again about 450 Myr after the first transit, its mass is entirely negligible.

\subsection{Model stability} \label{s:stab}

We have studied the long-term stability of our models with a series
of expensive simulations for an isolated galaxy run over 4 Gyr and at different resolutions. We made this investment because cold discs are susceptible to the emergence and evolution of spurious substructure (e.g. clumps, bars, spiral arms) as the disc evolves \citep{dOnghia2013,Fujii2018}. 
To underscore the importance of this test, the reader is encouraged to compare two disc simulations with N  $\approx 10^6$ and N $\approx 10^8$ particles
(see footnote~\ref{F:movie}). By $T\sim 1$ Gyr, the low-resolution
simulation excites complex bending modes in the outer disc that only get worse as time marches on, whereas no such effect is apparent in the high-resolution simulation.
In the high-resolution simulation, the surface density profiles are reasonably constant over the full timespan ($<$ 5 percent, rms), which amounts to about 20 disc rotations at the solar radius (Fig.~\ref{f:struct}, bottom right). The other panels show that deviations from the original rotation profiles are less than 1 percent at 
all radii. 

We also examined the stellar dispersion profiles in both $R$ and $z$.
The vertical dispersion profile $\sigma_z(R)$ is essentially unchanged at all radii ($<5$ percent), which shows that there is no disc heating induced by the coarseness of the finite resolution at any radius.
In separate tests, we find that we can avoid internal disc heating when the number of disc particles exceeds about 10 million particles. The radial dispersion profile $\sigma_R(R)$ shows a slight excess (13 percent on average) over 4 Gyr; we deem this to be acceptable given the goals of the current work.
From our perspective, the parametric profiles are essentially constant, such that the ultrathin disc retains the imposed ICs over many rotation periods, providing confirmation of the power of \agama\ in setting up robust models.

\begin{figure*}
\centering
\includegraphics[width=0.28\textwidth]{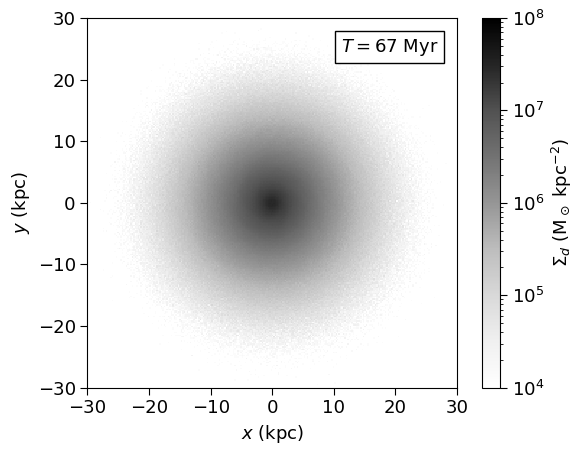}
\includegraphics[width=0.28\textwidth]{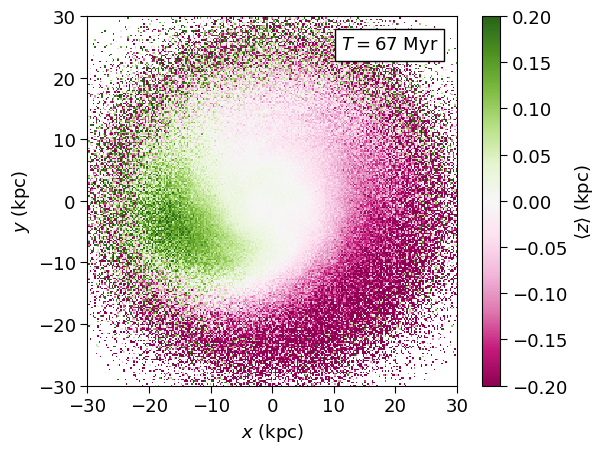}
\includegraphics[width=0.28\textwidth]{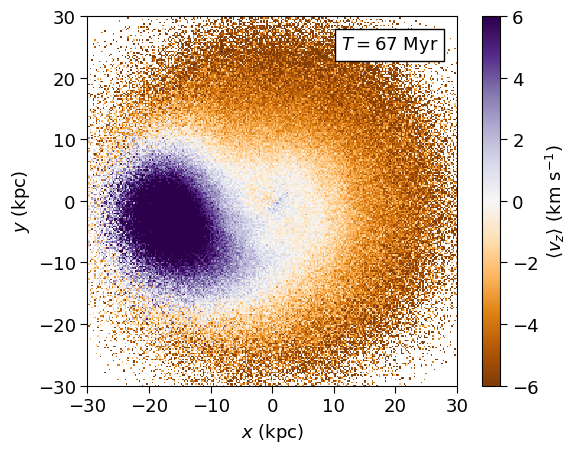}\\
\includegraphics[width=0.28\textwidth]{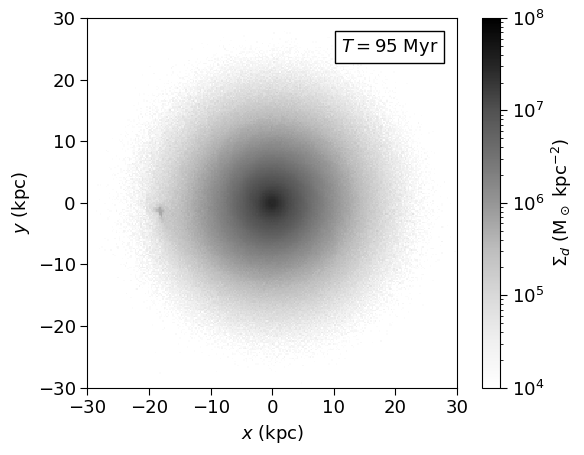}
\includegraphics[width=0.28\textwidth]{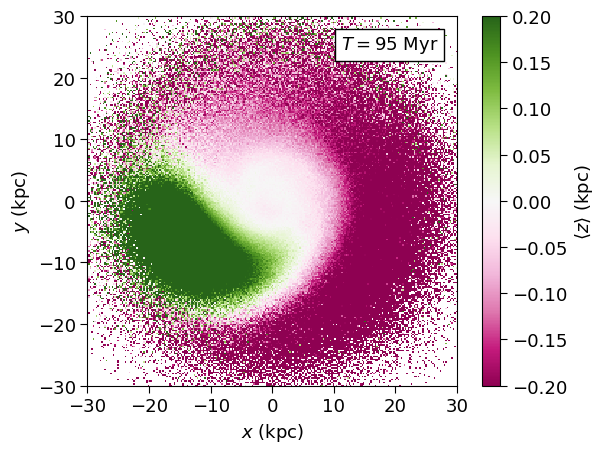}
\includegraphics[width=0.28\textwidth]{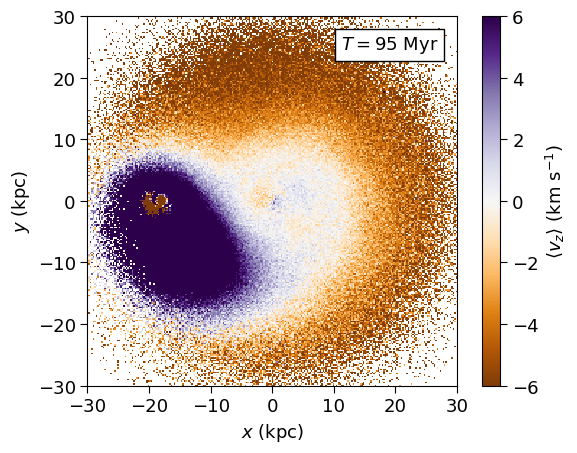}\\
\includegraphics[width=0.28\textwidth]{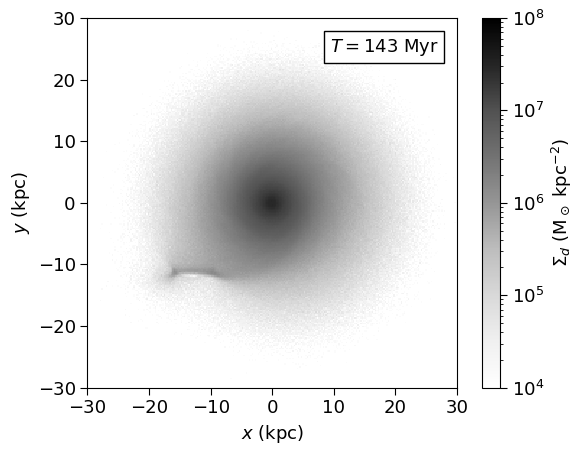}
\includegraphics[width=0.28\textwidth]{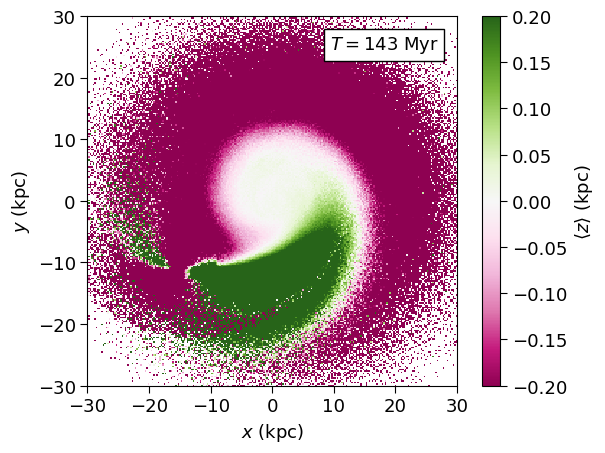}
\includegraphics[width=0.28\textwidth]{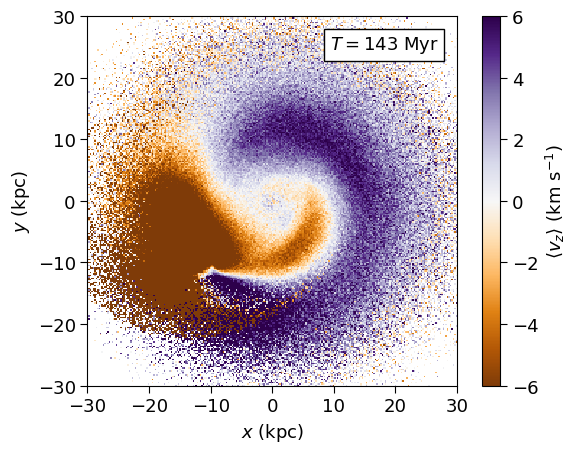}
\includegraphics[width=0.28\textwidth]{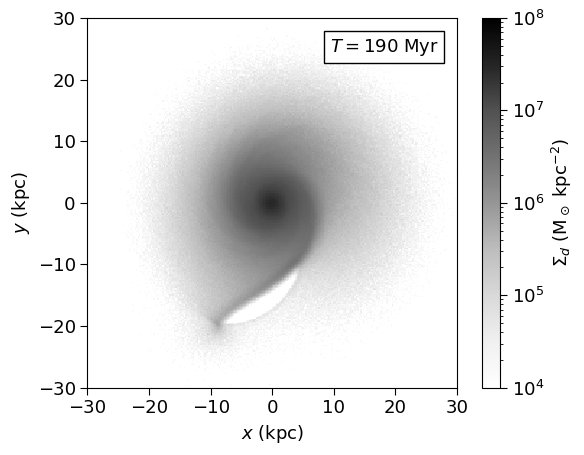}
\includegraphics[width=0.28\textwidth]{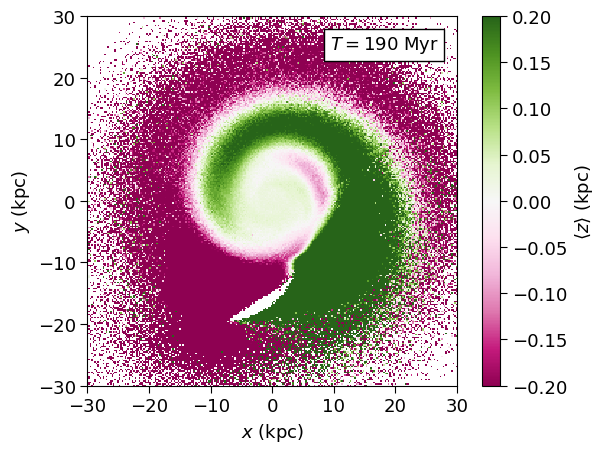}
\includegraphics[width=0.28\textwidth]{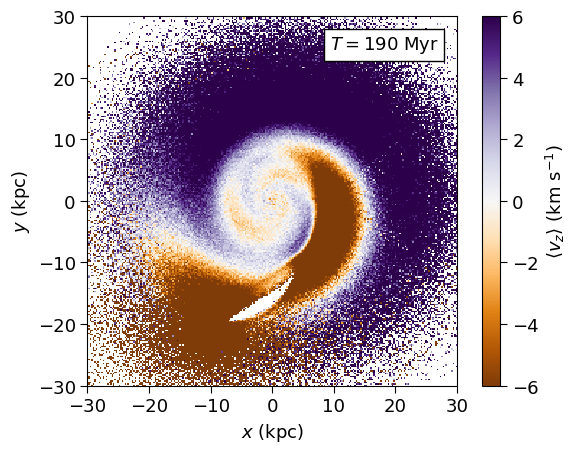}
\caption{The projected stellar density distribution of the disc (left), the average height of the stars (middle), and the average vertical speed of the stars (right) at four time steps ($T=67, 95, 143, 190$ Myr). The disc's rotation is counter-clockwise in this projection. The point mass crosses the disc at $(x,y) \approx (-18,0)$ kpc ($T=95$ Myr). The disturbance is already apparent at $T=67$ Myr when the perturber is about 20 kpc from the impact point. Stars in the vicinity of the impact point are initially lifted towards the intruder. A bending wave is triggered by the impulse and it immediately begins to wrap up with the differential rotation. {\it These images can be viewed as a movie sequence for the full 2~Gyr duration at the website indicated at footnote~\ref{F:movie}.}
}
\label{f:clock_xy}
\end{figure*}

\begin{figure*}
\centering
\includegraphics[width=0.28\textwidth]{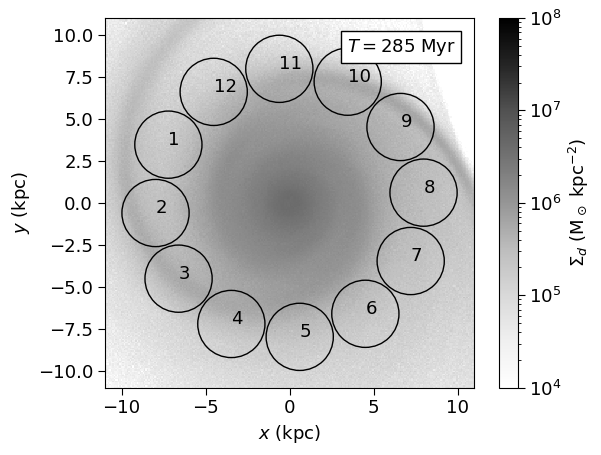}
\includegraphics[width=0.28\textwidth]{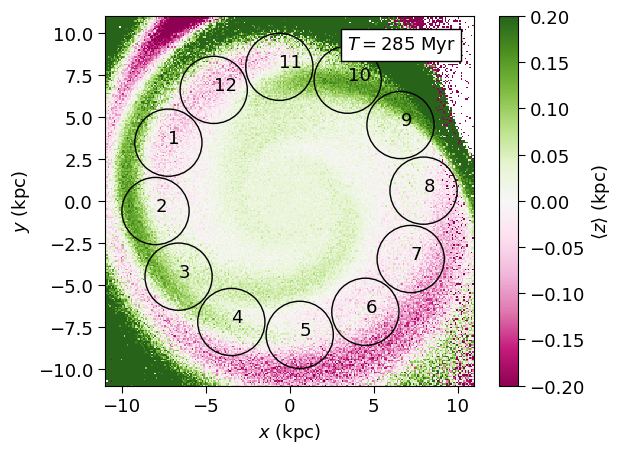}
\includegraphics[width=0.28\textwidth]{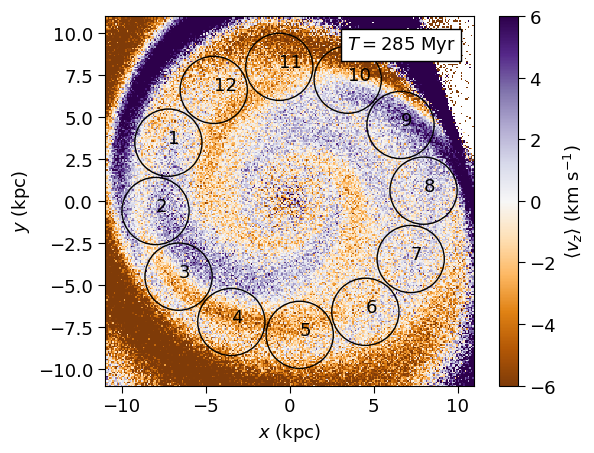}\\
\includegraphics[width=0.28\textwidth]{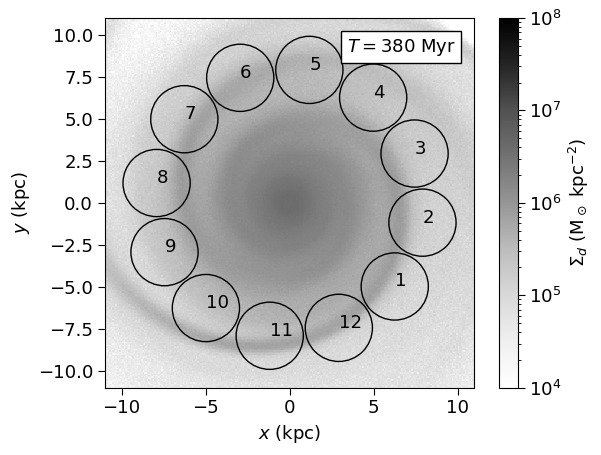}
\includegraphics[width=0.28\textwidth]{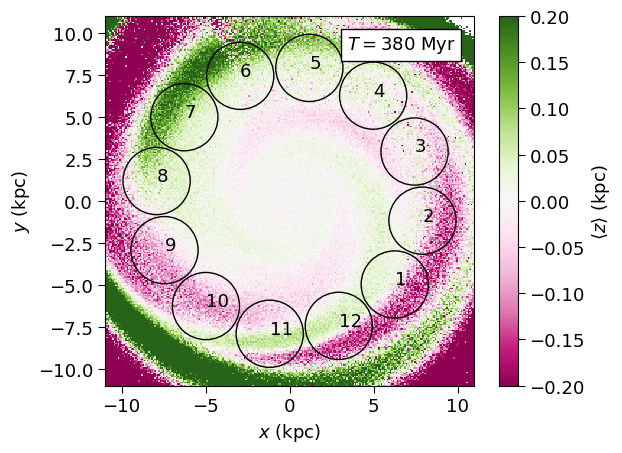}
\includegraphics[width=0.28\textwidth]{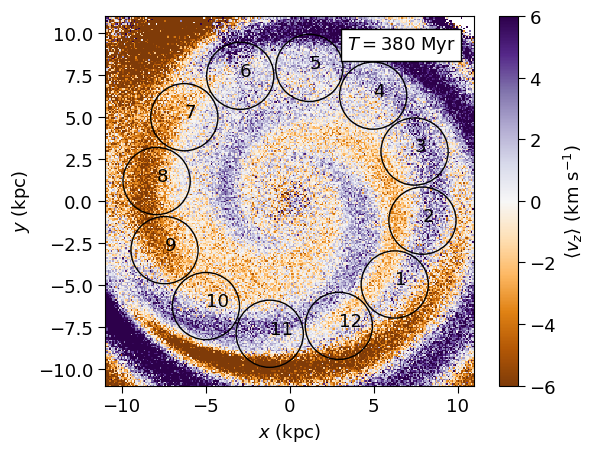}\\
\includegraphics[width=0.28\textwidth]{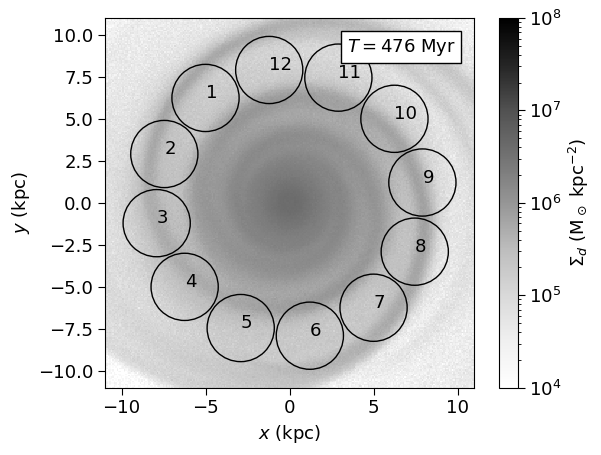}
\includegraphics[width=0.28\textwidth]{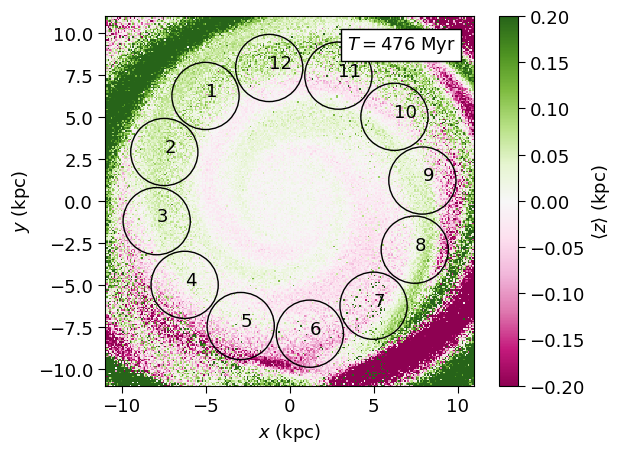}
\includegraphics[width=0.28\textwidth]{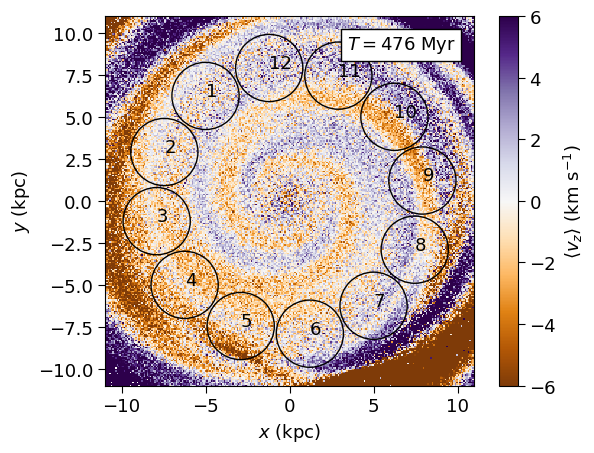}\\
\includegraphics[width=0.28\textwidth]{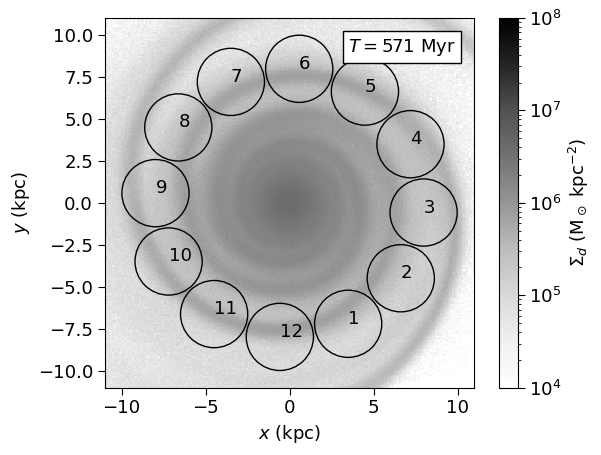}
\includegraphics[width=0.28\textwidth]{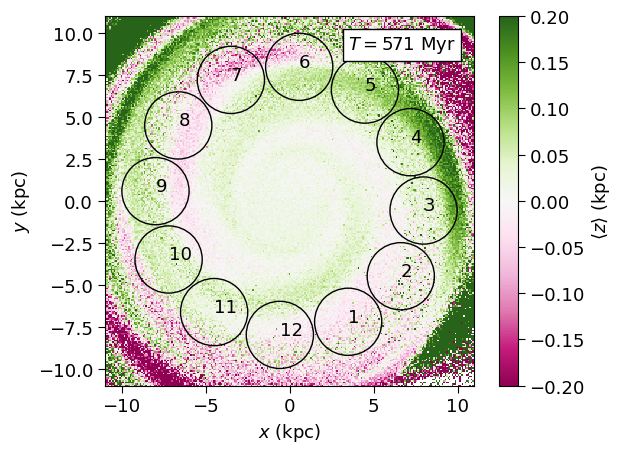}
\includegraphics[width=0.28\textwidth]{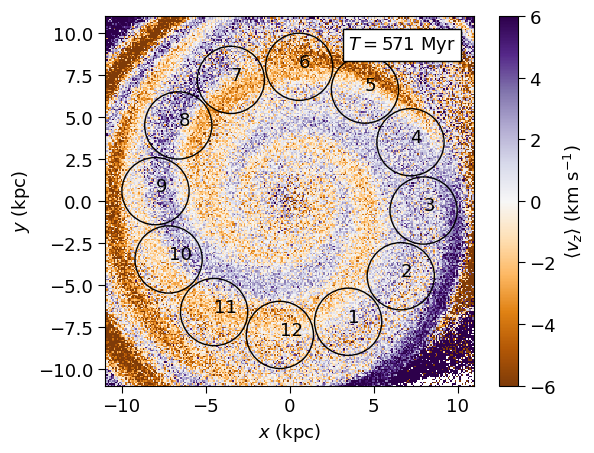}
\caption{(continued from previous figure) Early to middle stage evolution of the disc seen from the $z>0$ axis; the time sequence continues in the next figure. The first panel is the projected stellar density field in the $x-y$ plane. The second and third panels are the average vertical distance $\langle z \rangle$ and the average vertical velocity $\langle V_z \rangle$ respectively in the $x-y$ plane. The first panel shows the $m=2$ spiral density wave; the second and third panels reveal the $m=2$ bending mode. The numbered circles are 4 kpc in diameter and spread along the Solar Circle ($R=8.2$ kpc). These relate to our discussion of the phase spiral and show the rotation of the disc.
}
\label{f:clock_xy2}
\end{figure*}

\begin{figure*}
\centering
\includegraphics[width=0.28\textwidth]{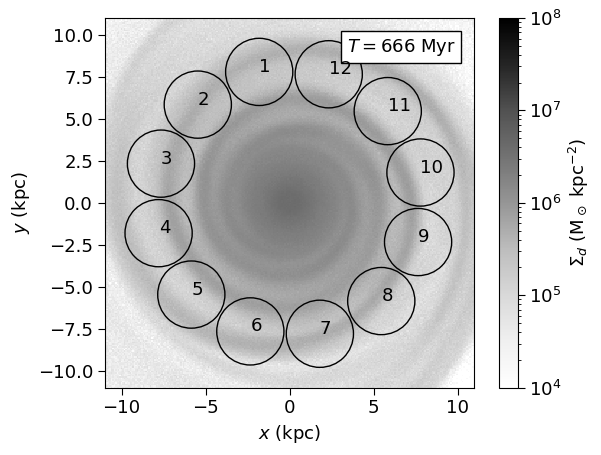}
\includegraphics[width=0.28\textwidth]{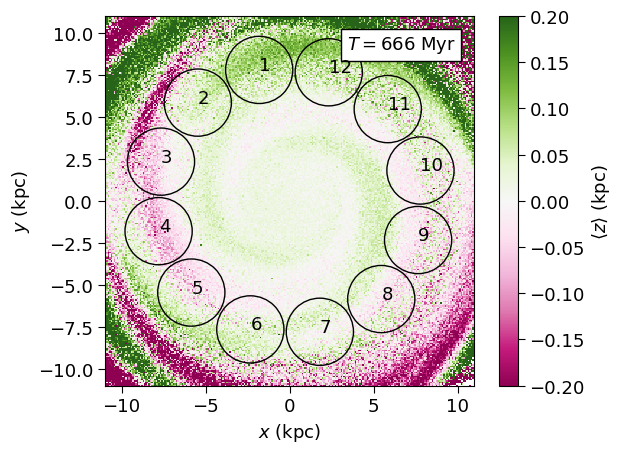}
\includegraphics[width=0.28\textwidth]{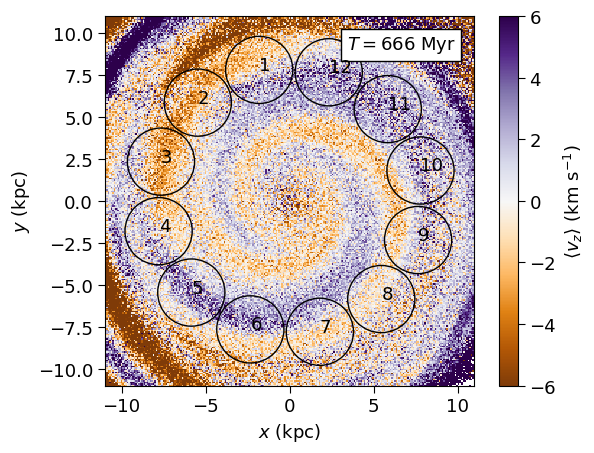}\\
\includegraphics[width=0.28\textwidth]{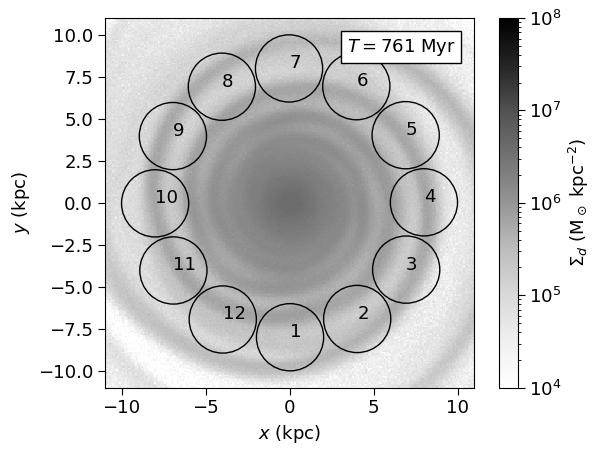}
\includegraphics[width=0.28\textwidth]{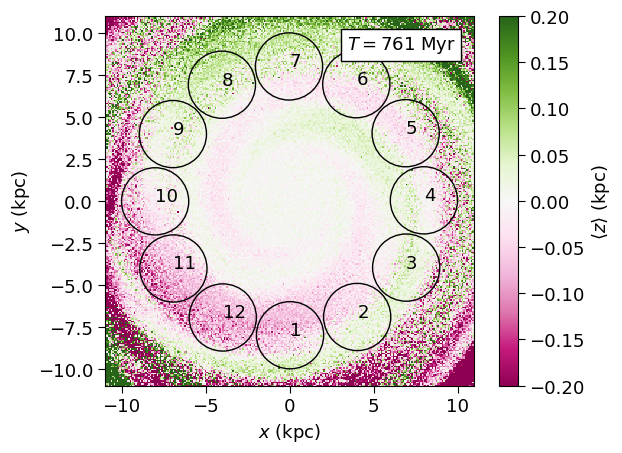}
\includegraphics[width=0.28\textwidth]{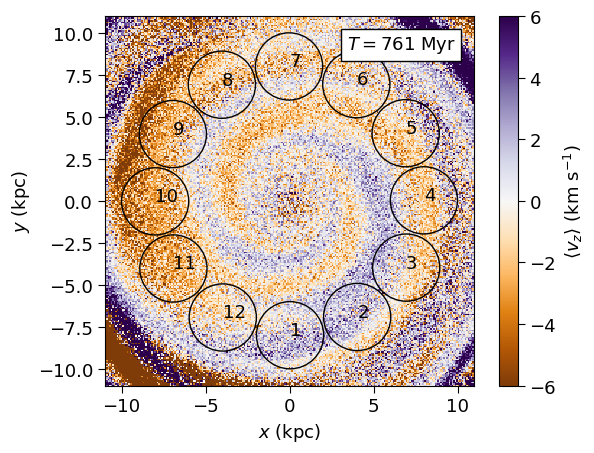}\\
\includegraphics[width=0.28\textwidth]{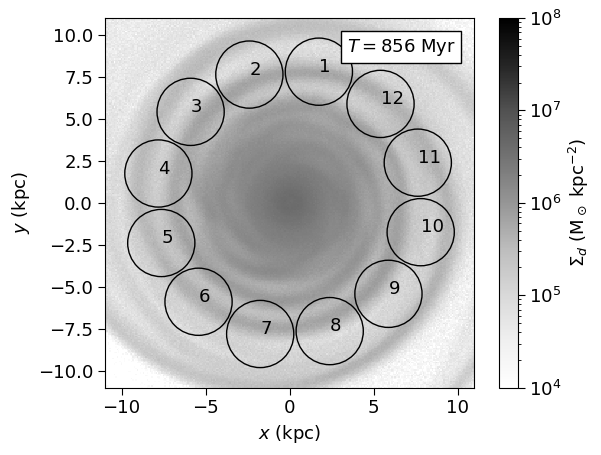}
\includegraphics[width=0.28\textwidth]{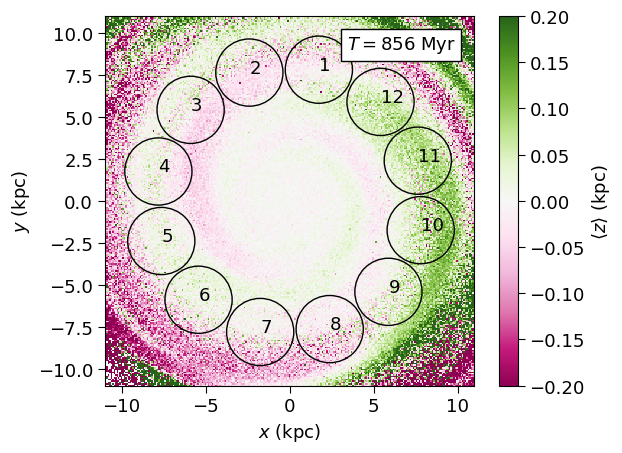}
\includegraphics[width=0.28\textwidth]{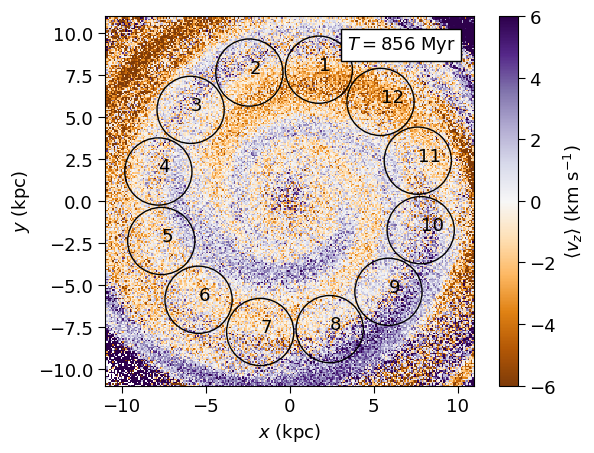}\\
\includegraphics[width=0.28\textwidth]{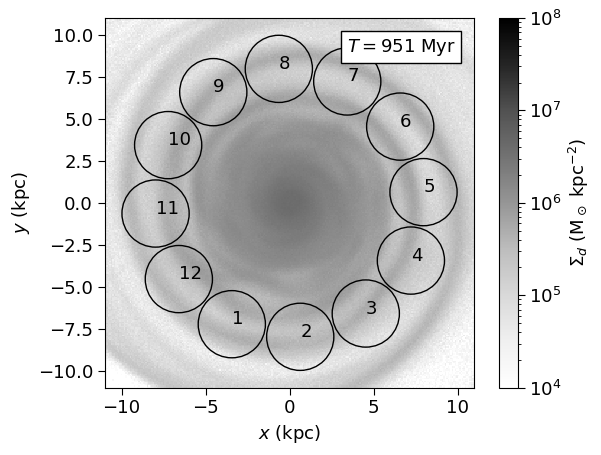}
\includegraphics[width=0.28\textwidth]{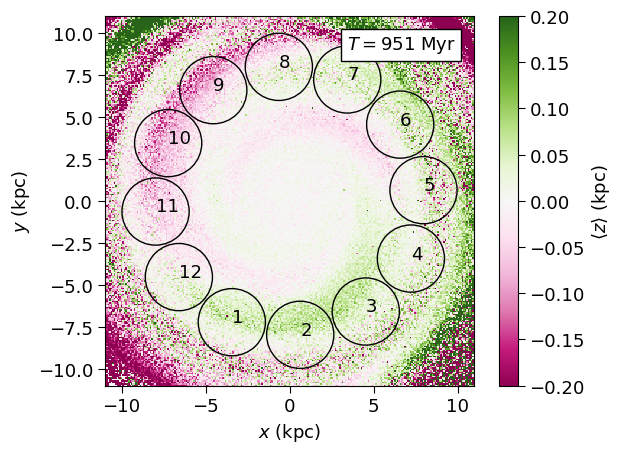}
\includegraphics[width=0.28\textwidth]{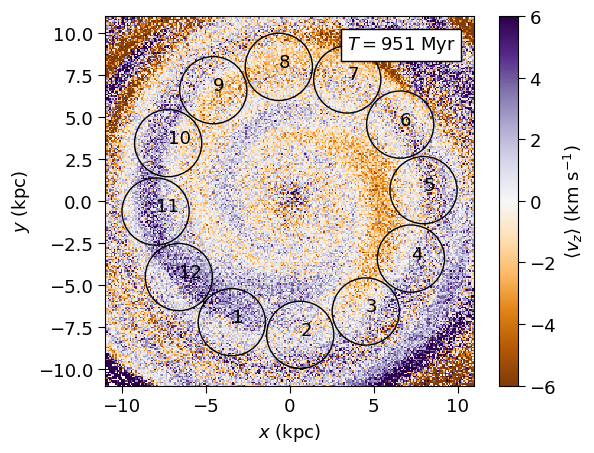}
\caption{(continued from previous figure) Middle stage to late stage evolution of the disc seen from the $z>0$ axis.}
\end{figure*}

\begin{figure*}
\centering
\includegraphics[width=0.7\textwidth]{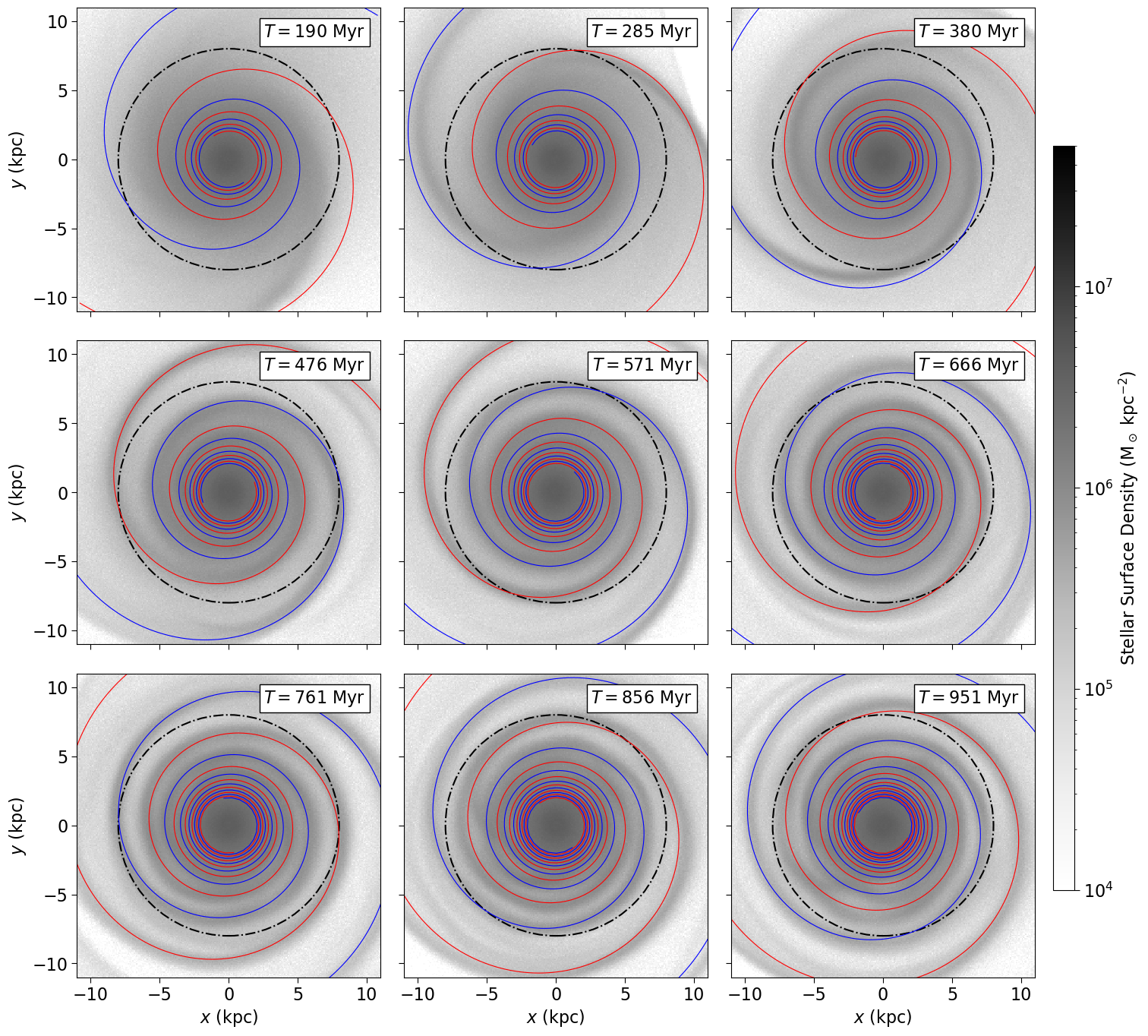}
\caption{Middle and late stage evolution of the kinematic density wave in the N-body simulation. The rotating galaxy is viewed at each time step from an inertial (non-rotating) frame. The solar circle is indicated with the dot-dashed line. After the phase of the pattern is fixed at $T=571$ Myr, there are no free parameters for the functional form of the spiral model in the other time steps. For 200 Myr after the impulse, the non-symmetric spiral pattern is evolving towards 2-fold symmetry. After $T=300$ Myr, the bisymmetry is revealed and the correspondence is very good, although a vestigial non-bisymmetry exists at all time steps. The pattern is winding up as $\phi(R,t) \propto (\Omega(R)-\frac{1}{2}\kappa(R))\;t$. This is the natural response of a cold disc perturbed by a strong impulse.
}
\label{f:lindblad}
\end{figure*}

\begin{figure*}
\centering
\includegraphics[width=0.45\textwidth]{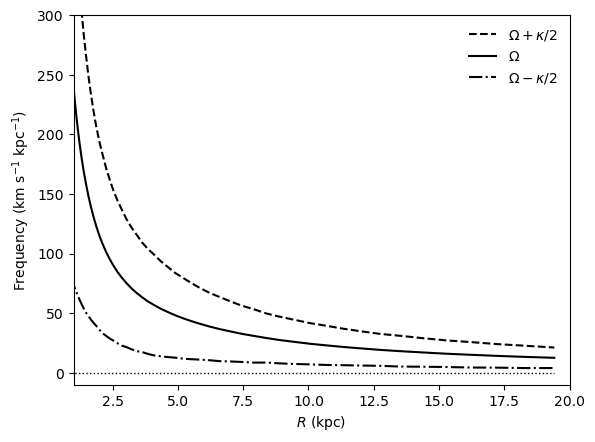}
\includegraphics[width=0.45\textwidth]{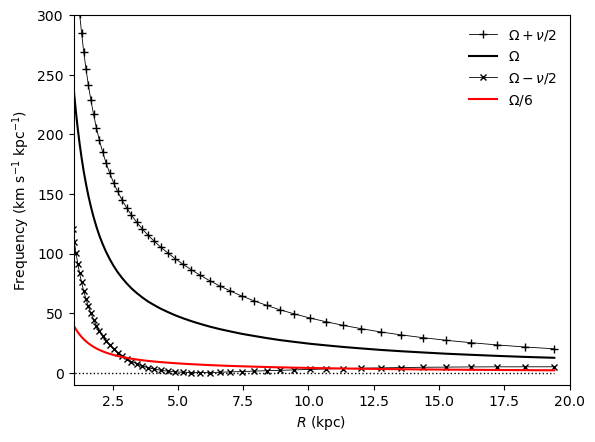}
\caption{(Left) The angular frequency ($\Omega$) and the planar Lindblad resonance ($\Omega-\kappa/2$) as a function of radius $R$ measured from the isolated disc simulation in the mid-plane ($z=0$). (Right) The angular frequency and vertical Lindblad resonance ($\Omega-\nu/2$) measured from the same simulation in the mid-plane ($z=0$). Note that $\Omega$ (solid curve) is identical in both panels.
}
\label{f:freq}
\end{figure*}

\section{Analysis and results} \label{s:analy}

\subsection{Early stage evolution} \label{s:early}

In analysing the simulations, our first goal is to understand just how the cold disc experiences the impulsive force from the disc-crossing intruder. {\bf A key consideration} is to determine how efficiently the perturber's impulse is able to trigger a response in the cold disc.
This must depend both on the infall velocity ($\vec{v}_{\rm P}$) and the perturber mass ($M_{\rm P}$), with lower velocities and higher masses triggering a stronger response. 
Earlier simulations \citep[][]{Darling2019b} employ perfect coupling by inserting the bending modes by hand. {\bf Our second goal is} to understand how the phase-spiral signal builds over time as a diagnostic for age-dating the phenomenon. {\bf This is crucial for understanding which disc transit was responsible for the corrugated disc we observe today. The answer to this question will tell us a great deal about the mass loss rate along Sgr's orbit.}

First, we investigate how well the BS18 picture is reproduced in our simulations. Movies of the disc-perturber interaction are available at our website; see Footnote~\ref{F:movie}. We stress that, at each timestep, all results are shown in the reference frame of the simulated galaxy's centre of mass where the disc's angular momentum vector is aligned with the positive $z-$axis. The perturber moves parallel to the $+z$ axis towards the plane.

In Fig.~\ref{f:clock_xy}, the
response of the disc is presented in the $x-y$ plane at $T=67$ Myr, $T=95$ Myr (time of transit), $T=143$ Myr and $T=190$ Myr. {\bf For all figures that follow, only the disc particles are shown.} Each time step shows three panels: (1) the projected stellar density; (2) the projected mean $z$-height $\langle z\rangle$ of the stars; (3) the projected mean $z$-motions $\langle V_z\rangle$ of the stars.
In the $x-z$ projection, even before the disc crossing ($T=67$ Myr), stars move in $z$ towards the perturber along the impact trajectory. These are mostly stars that were moving {\it away} ($\phi-\phi_o > 0$) from the impact point ($\phi_o=0$) in the disc plane (cf. BS18, Fig. 1) after transit ($T> 95$ Myr). 

Conversely, after the perturber has transited the disc, stars are observed to pursue the perturber along its impact trajectory. These are mostly stars that were moving {\it towards} ($\phi-\phi_o < 0$) the impact point in the disc plane at the time of transit. This picture is in good agreement with the BS18 model, although it misses an important ingredient, as explained below.

The early phase of the disc's evolution is broadly consistent with the BS18 picture summarised in Sec.~\ref{s:BS18}, except they do not consider the influence of the shearing disc due to differential rotation. We believe this to be crucial to how the phase spiral emerges. At $T=143-190$ Myr, the first signs of a density wave become evident in the projected density map. Moreover, the impulse sets up a strong, low-order ($m=1$) bending mode across the disc seen initially as the sine-wave warp of the outer disc in the $\langle z\rangle$ and $\langle V_z\rangle$ panels. Of particular interest is the transition from $m=1$ in the outer disc to an $m=2$ bending mode inside $R\approx 10$ kpc that wraps up with the differential rotation. (We explore this behaviour with a model in a later section.)

The early stages set up a sequence of events that unfold in the remaining time steps for which we show representative images sampled at intervals of 95 Myr (Fig.~\ref{f:clock_xy2}). The perturber triggers both a density wave (left panels) and a bending wave (middle and right panels) that wrap up as time passes.
The panels are encoded in the same way as Fig.~\ref{f:clock_xy} but now zoomed to show the inner disc ($R<11$ kpc) more clearly. Note that the vertical motion $\langle V_z \rangle$ is out of phase with the vertical displacement $\langle z\rangle$. This is expected for a corrugated wave (bending mode) that oscillates up and down as it wraps up. Observe also that the $\langle z\rangle$ displacement increases with radius as expected (B19, Figs. 20 and 21). 

The twelve circles shown in each of the panels of Fig.~\ref{f:clock_xy2} are the sampled volumes in our later discussion of phase-spiral evolution. Note three things: (i) our solar neighbourhood is ``volume 1" ($x=-8.2$ kpc) and the intruder crossing point is at $x=-18.0$ kpc; (ii) at $T=95$ Myr, these lie along the same radius vector at disc transit; (iii) at $T\approx 200$ Myr, volume 1 has moved to the other side of the disc since the impact, consistent with the orbital period at the Sun's radius ($T_\odot \approx 210$ Myr), while the crossing point ($T_C \approx 460$ Myr orbital period) has only moved by about 80$^\circ$ bringing it in line with volume 10. After a full orbit of volume 1 ($T=305$ Myr in our time frame), the crossing point now lies along the radius vector with volume 7.

\subsection{Middle to late stage evolution}

\subsubsection{Kinematic density wave}

For eighty years, we have known that galaxy interactions excite trailing spiral arms in disc galaxies \citep{Holmberg1941}. Thereafter, Lindblad discovered that the force fields of disc galaxies preserve twofold symmetric (bisymmetric) structures against dissolution from the differential rotation \citep{Lindblad1956,Lindblad1960}.
In an isolated galaxy, stellar orbits are mostly elliptic and precess about the centre. An external perturber can organize the elliptic orbits into a coordinated precession (2-arm spiral pattern) that depends on the local angular frequency $\Omega(R)$ and radial frequency $\kappa(R)$, subject to the shearing disc's lowest order resonances, $\Omega(R)\pm\kappa(R)/2$ \citep{Kalnajs1973}.
These features are observed in numerical simulations to be associated with density waves that wind up slowly, rather than material arms that wind up rapidly with the disc's differential rotation \citep[][]{Sundelius1987,Oh2008,Dobbs2014}.

The spiral pattern is the natural response of a cold, differentially rotating disc subjected to a strong impulsive force \citep[][\S6.2]{Binney2008}. For an $m=2$ density wave with trailing arms ($\Omega > \kappa/2$), as seen from the positive $z$ axis, the spiral pattern $\phi_D(R)$ ($R$ in units of kpc) winds up following
\begin{equation}
    \phi_D(R,T) = \left(\frac{\Omega_D(R)+\Omega_{\rm o}}{1\; \rm km\; s^{-1}\; kpc^{-1}}\right) \left(\frac{T}{978 \;\rm Myr}\right) + \phi_{D,{\rm o}}
    \label{e:lindblad1}
\end{equation}
where
\begin{equation}
    \Omega_D(R) = \Omega(R)-\frac{1}{2}\kappa(R)
        \label{e:lindblad2}
\end{equation}
and $\Omega_{\rm o}$ is included in the fitting process to account for any pattern speed due to figure rotation. ($\phi_{D,{\rm o}}$ is an arbitrary offset.)
All frequencies are in units of \kms\ kpc$^{-1}$ (equivalent to $(978{\rm \; Myr})^{-1}$), $T$ is in units of Myr, and angles are in radians. 

If $\Omega_{\rm o}=0$, this form describes a ``kinematic" density wave because it depends only on the kinematics of elliptic orbit crowding at aphelia. In this instance, the
spiral pattern is not a propagating wave mode, it has no group velocity, and the complicating effects of self-gravity and swing amplification are negligible. For a flat rotation curve, a kinematic density wave wraps up at about 30\% of the angular rate of a material arm carried by the shearing disc ($\Omega_D = \Omega$).

Interestingly, this simple form is not exactly what is observed in simulations of disc mergers and perturbed discs to date \citep{Struck2011}. For example, the ``grand design'' spiral in M51 is a common target for modellers. Numerical simulations of the 2-arm spiral pattern are found to wind up at a higher rate than indicated by Eq.~\ref{e:lindblad1}, with a non-linear dependence on radius \citep{Oh2008,Dobbs2010,Salo2000}.

So what do we find? In the last section, we saw how the density wave developed in the projected stellar density map (Fig.~\ref{f:clock_xy2}).
For the wave to be kinematic, it must simply wind up with no additional contribution from, say, figure rotation with an associated pattern speed.
In Fig.~\ref{f:lindblad}, the model described by Eq.~\ref{e:lindblad1} is overlaid on the density map where both the arm (blue: $\phi_D$) and counter arm (red: $\phi_D+\pi$) are shown. The parameters $\Omega(R)$ and $\kappa(R)$ are measured from the isolated disc simulation and presented in Fig.~\ref{f:freq} (left panel). After determining $\Omega = V_\phi/R$, we calculate
\begin{equation}
    \kappa^2(R) = \left(\frac{\partial^2\Phi_{\rm eff}}{ \partial R^2}\right)_{z=0} =
    \left( R\frac{\d\Omega^2}{dR} + 4\Omega^2\right)_{z=0}
    \label{e:kappa}
\end{equation}
where $\Phi(R)_{\rm eff}$ is the effective gravitational potential taken from the simulation, and the subscript $z=0$ indicates that all measurements are made in the plane. In the early stages, at the time of the disc transit, the spiral pattern is highly asymmetric, but the pattern clearly settles down after $T=190$ Myr and starts to exhibit 2-fold symmetry. After $T=450$ Myr, the Lindblad-Kalnajs pattern (Fig.~\ref{f:lindblad}) reveals itself and tracks the evolution of the spiral pattern for the remainder of the simulation.

Once the functional form of the pattern is fixed, there are no free parameters for any of the other time steps shown. But this statement requires two qualifications: (i) The time variable $T$ has an arbitrary offset $T^\prime$ that determines the degree of winding when the spiral pattern first appears.
An initial line of points is assumed to lie along $y>0$ and $x=0$; the points are then advanced in time to match the spiral pattern at some epoch \citep{Binney2008}. For our model to match, we set $T^\prime = 550$ Myr. (ii) Intriguingly, we find that the spiral pattern does have some figure rotation. We measure this to be $\pi/24$ rad every 100 Myr once the density wave has settled down from the initial impulse. Using the conversion factor above, the rigid pattern speed is equivalent to $\Omega_{\rm o} = 1.3$ \kms\ $\kpc^{-1}$. While $\Omega_{\rm o}$ is very slow, it is non-zero and indicates that the spiral arms are propagating as a dynamical wave mode, rather than a kinematic anomaly \citep[cf.][]{Lin1966}.

In summary, the evolution of the $x-y$ spiral has two components: (a) a wind-up rate given by the simple Lindblad-Kalnajs formula (eq.~\ref{e:lindblad2}); (b) an additional slow figure rotation $\Omega_{\rm o}$ that rotates the fixed pattern in a prograde (anticlockwise) direction as one moves forward in time. The combined effect is much slower than the wind-up rate of material arms.

As far as we can establish, the most basic form for the spiral evolution described in Eq.~\ref{e:lindblad1} has not strictly been observed before in simulations, certainly not over the long time frames considered here. This may reflect the high level of self-consistency made possible by the \agama\ initialisation of our cold disc simulations. Furthermore, it appears to be a {\it travelling} density wave $-$ the `Lindblad-Kalnajs' propagating wave mode.

\subsubsection{Kinematic bending wave}

\begin{figure*}
\centering
\includegraphics[width=0.9\textwidth]{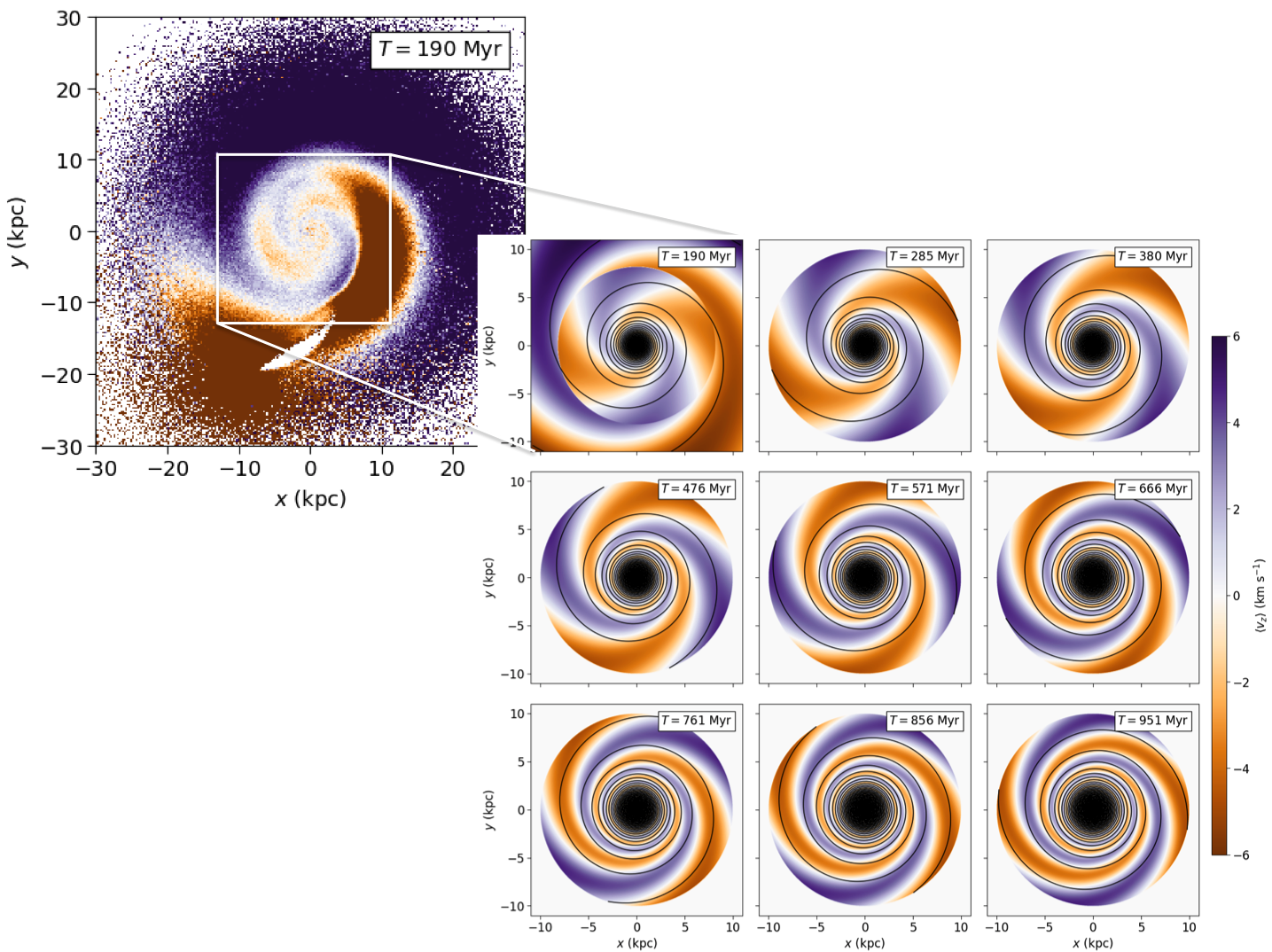}
\caption{A comparison of the evolution of the bending wave (colour image) with the wrapping up of the kinematic density wave (spiral lines) for $R \leq 10$ kpc. The colour image represents the mean vertical velocity field $\langle v_z \rangle$ from $-6$ \kms\ (blue) to $+6$ \kms (red). The interplay of the two distinct wave modes wrapping up at different rates is the origin of the phase spiral, as we discuss. Both are model fits to the disc simulations. At the first time step ($T=190$ Myr), we include a composite model that is to be compared to the inset. This shows how the $m=1$ bending mode ($R>8$ kpc) relates to the inner $m=2$ bending mode ($R<8$ kpc).}
\label{f:wrap}
\end{figure*}

In addition to in-plane spiral density waves, corrugated waves are now well established in the Galaxy (Sec.~\ref{s:intro}). How these work together, if at all, is entirely unclear. A bending wave is observed in our new simulation, and appears to wrap up like the density wave discussed in the last section. They appear to share the same common origin, but are these a superposition of modes that propagate without interference from the other? Like spiral density waves, bending modes may be transient on a timescale of a few rotation periods at the Sun's radius \citep{Hunter1969,Sparke1984,Binney1998}. 

The progress that has been made in understanding spiral density waves is restricted to 2D razor-thin discs. Just how this action extends to 3D discs is largely unknown \citep{FouvryPichon2017}. It is clear
that the fundamentally in-plane mode must involve $V_z$ in addition to
$V_\phi$ because stars will be drawn down to regions of high density \citep{Masset1997}. These motions will remain conjectural until the theory of spiral structure has been extended from razor-thin discs, in which vertical motion is impossible, to discs of finite thickness, a 
very difficult proposition \citep{Binney2008}.

The available formalism relating to the second kind of mode, the corrugation or bending wave, is even more primitive than the current theory of spiral structure because it involves neglecting epicyclic oscillations in addition to taking the disc to be razor
thin \citep{Hunter1969}. Hence we really have very little idea what a
proper theory of corrugation waves would look like. It is plausible
that the $V_\phi$ motions are coupled with $V_z$ motions because warps (that can be considered as concentric rings) arise from torques exerted by one ring on another.

Both wave phenomena are defined by natural oscillation frequencies and are described by dispersion relations that look similar, although the dependence on the disc self-gravity has the opposite sign \citep{Binney2008}.
The vertical frequency has a similar form to the epicyclic frequency in Eq.~\ref{e:kappa}, viz.
\begin{equation}
    \nu^2(R) = \left(\frac{\partial^2\Phi_{\rm eff}}{ \partial z^2}\right)_{z=0} 
    \label{e:nu}
\end{equation}
The vertical height of any bending mode with parameter $m$ is described by \citep[e.g.][]{Binney2008}
\begin{equation}
    z_B(R,\phi,T) = z_o(R)\cos[m(\phi-\Omega_B T)]\cos\nu T
    \label{e:bend}
\end{equation}
The kinematic bending modes are subject to winding up much like the density waves. The pattern speed of the bending wave is
\begin{equation}
    \Omega_B = \Omega(R) \pm \nu(R)/m
\end{equation}
We now investigate the bending mode in our simulation because of its role in the phase-spiral phenomenon discussed below. First, we measure $\nu(R)$ directly from the isolated disc simulation using Eq.~\ref{e:nu}; the results are presented in Fig.~\ref{f:freq}.

Our expectation is that the winding-up of the $m=2$ bending model will lag behind the $m=2$ spiral density wave. In a spherical gravitational potential with average density $\rho_s$, $\kappa=\nu$; for a flattened disc potential with density $\rho_d$, $\nu^2/\kappa^2 \approx (3/2) ~(\rho_d/\rho_s)$. For example, along the solar circle, $\nu/\kappa\approx 2$ \citep{Binney2008}. Thus we expect
$\Omega - \nu/2 < \Omega - \kappa/2$ and this is indeed what we find when comparing both panels in Fig.~\ref{f:freq}.
Fig.~\ref{f:clock_xy2} presents a comparison of the projected stellar density, the average vertical height $\langle z \rangle$ and the average vertical velocity $\langle V_z \rangle$ in the $x-y$ plane. The slow wrapping up of the bending wave through the oscillatory behaviour in $\langle z \rangle$ and $\langle V_z \rangle$ is clearly evident in the last two columns. 

We now model these patterns using Eq.~\ref{e:bend} and the measured properties of the disc (Fig.~\ref{f:freq}). The spiral patterns in $\langle z \rangle$ and $\langle V_z \rangle$ are out of phase by $\pi/2$ consistent with their oscillatory motion. We present our toy model for $\langle V_z \rangle$ in Fig.~\ref{f:wrap} to be compared to the last column in Fig.~\ref{f:clock_xy2}. The early time step is included to illustrate the important transition that was discussed in Sec.~\ref{s:early}. The $m=1$ integral sign warp is seen early on, but a stronger $m=2$ mode develops across the inner disc by $T=190$ Myr and dominates the later evolution.
After this time, the outer disc behaviour is complex, but the inner disc settles down to a well organized $m=2$ bending mode that wraps up slowly.

Our model for the $m=2$ bending wave is not as trivial as the form used for the spiral density wave. The equivalent form would be $\Omega_B = \Omega(R) - \nu(R)/2$ but this is
at odds with what we find.
In Fig.~\ref{f:freq}, note that $\Omega(R) - \nu(R)/2\approx 0$ in the range $4 \lesssim R \lesssim 6$ kpc. But the simulated disc continually winds up for all time over this radial range. Other values of $m > 2$ wrap up too rapidly. In its place, we use a slightly modified form, i.e. 
\begin{equation}
    \Omega_B={\rm max}[\:\Omega(R) - \nu(R)/2,~\Omega(R)/6\:]
\end{equation}
such that the measured trend is used in the inner disc, but beyond $R>3$ kpc, the angular frequency transitions to $\Omega_B=\Omega/6$ to ensure the `drop out' near $R=5$ kpc is never reached (red curve in Fig.~\ref{f:freq}). Our choice of this minimum threshold can be defended. The transient density wave model is defined by $\Omega_D \approx \Omega/3$; the bending mode must be slower and therefore we adopt $\Omega_B \approx \Omega_D/2$ after some experimentation.
This approximation tracks the observed wrapping up of the bending mode quite well. Because of our assumption, we are not able to detect reliably any fixed pattern speed by analogy with the density wave, and therefore we do not fit for it.

In Fig.~\ref{f:wrap}, the model for the evolving spiral density wave is superimposed on the bending wave in each time step. There is clearly a complex interaction between both wave modes as they wrap up at different rates. {\it We return to this behaviour below as it is central to our explanation for the phase spiral phenomenon.}

\begin{figure*}
\centering
\includegraphics[width=0.52\textwidth]{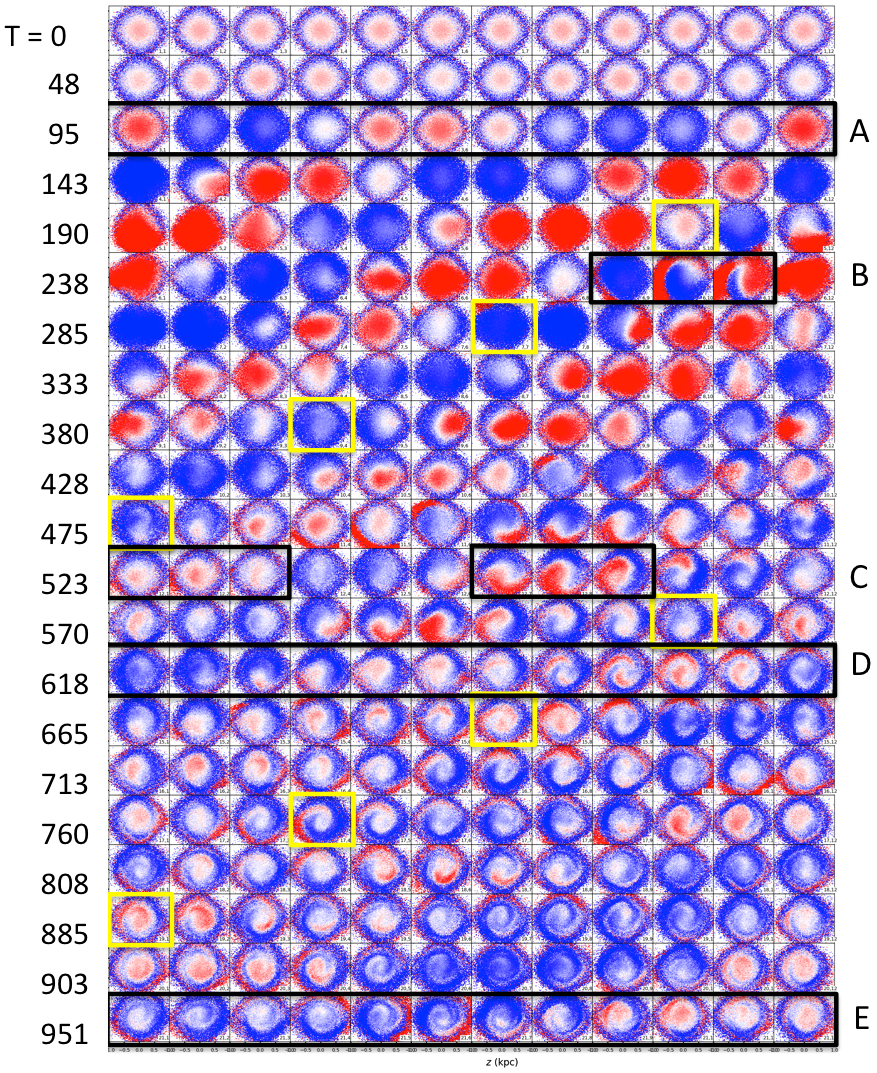}
\includegraphics[width=0.46\textwidth]{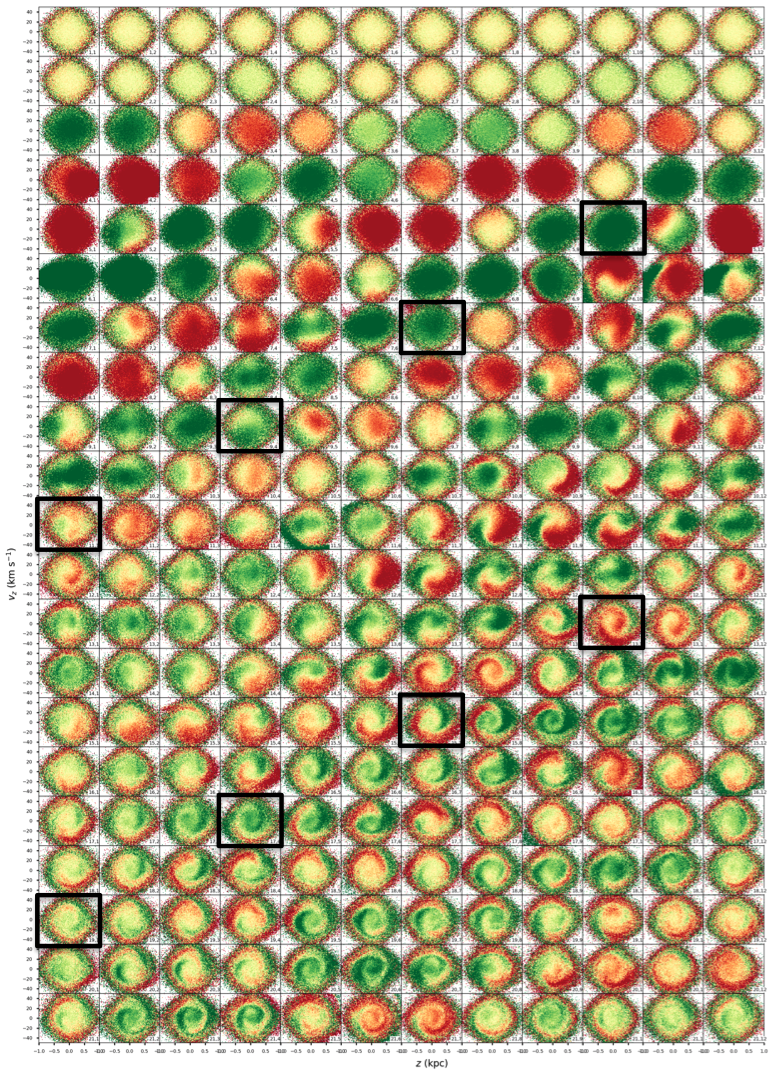}
\caption{Along each row, the $i$th volume in Fig.~\ref{f:clock_xy2} corresponds to the $i$th column. Each column corresponds to a fixed volume moving with the rotation: the solar neighbourhood is in column 1; its antipode is in column 7. Each row corresponds to a time step separated by 47.5 Myr as indicated. Moving to the right along a row corresponds to moving counter-clockwise by 30$^\circ$ per column along the solar circle, and the figure wraps around from the last to the first column on the left. The colour coding indicates the mean value of $\langle V_\Phi\rangle$ (left) and $\langle V_R\rangle$ (right); the ranges are (230, 250) \kms\  and (-10,+10) \kms\ respectively.
Regions A-E are discussed in the paper. The yellow (left) and black (right) boxes indicate the volumes that align with the original impact site, which lags further behind the solar neighbourhood as time passes. Higher resolution images are available at the website {\bf along with a complete time sequence to 2 Gyr} (footnote~\ref{F:movie}).}
\label{f:PScollage}
\end{figure*}

\begin{figure*}
\centering
\includegraphics[width=0.6\textwidth]{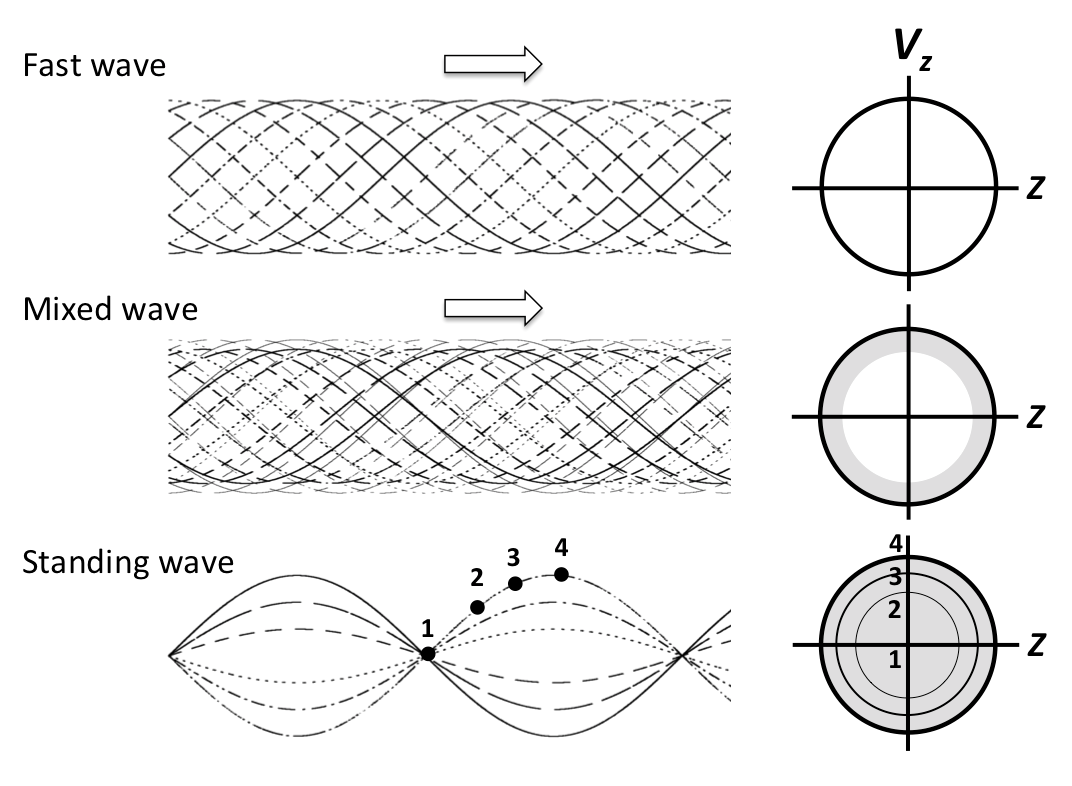}
\caption{Examples of how the $z-V_z$ plane can be filled by three distinct bending wave modes: a fast wave with constant amplitude, a composite wave with mixed amplitudes, and a standing wave. To the right, the idealized $z-V_z$ phase plane is shown. The set of wave modes operating within a volume determine how the $z-V_z$ plane is filled.}
\label{f:PSwaves}
\end{figure*}

\begin{figure*}
\centering
\includegraphics[width=0.85\textwidth]{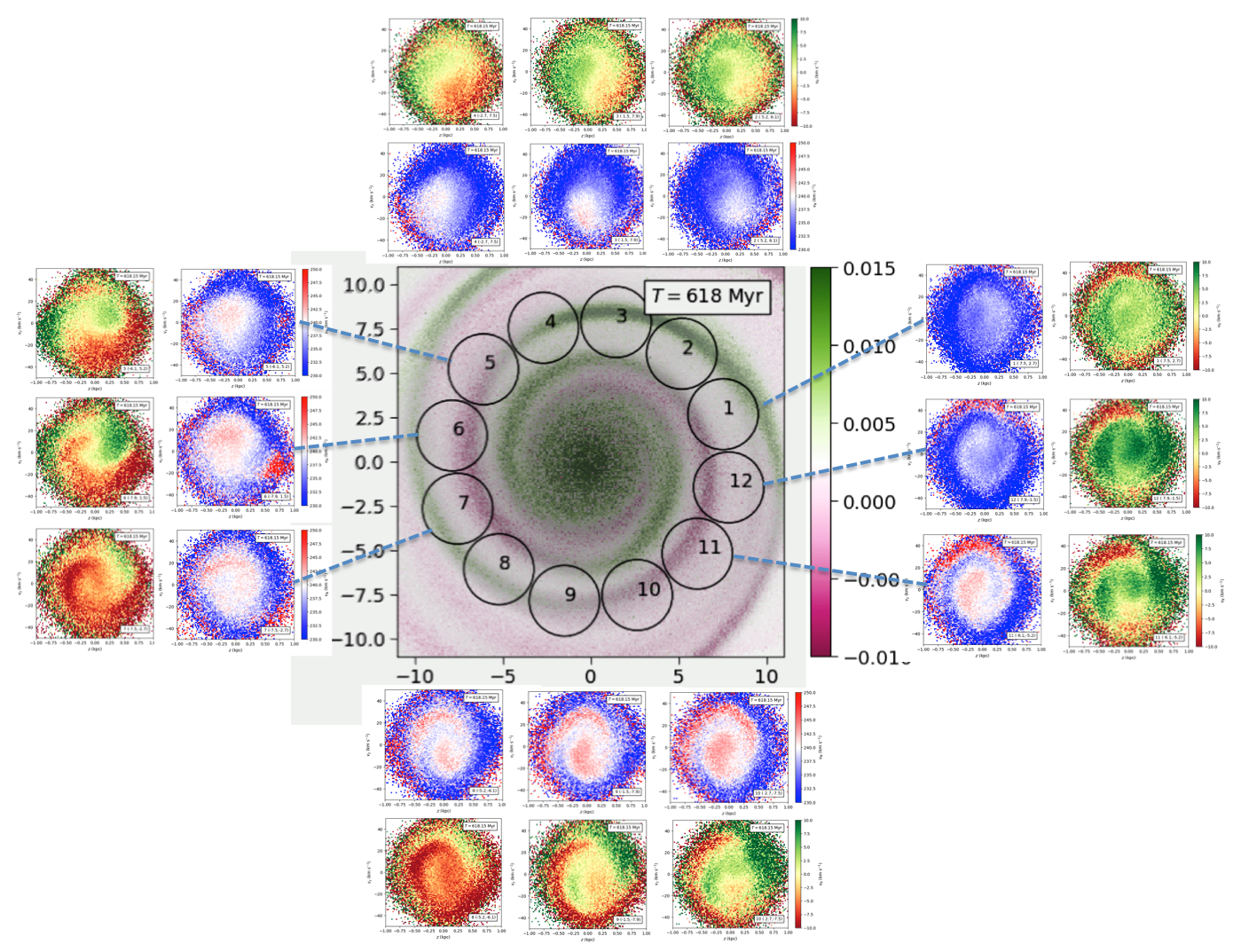}
\caption{A density-weighted $\langle z\rangle$ map at time step $T=618$ Myr. In the central $11\times 11$ kpc panel, the colour coding shows $\langle z\rangle$, the mean height (kpc) of the stars above and below the plane. The vertical undulations arising from the bending mode are clearly seen. Individual spiral arms roll up and down as the bending mode moves underneath them. The pink-coloured inner arms passing through volumes 7 and 12 curve inwards avoiding the next volume; note how the bending mode under both inner arms crosses over to the counter arm. The outer collages for $\langle V_\phi\rangle(z,V_z)$ and $\langle V_R\rangle(z,V_z)$ are taken from Fig.~\ref{f:PScollage}; the ranges are (230, 250) \kms\  and (-10,+10) \kms\ respectively.}
\label{f:bend-dens}
\end{figure*}

\subsection{Phase spiral evolution}
\label{s:PSevol}

We now describe the origin and evolution of the phase spiral as observed in our impulse-driven simulation. In Fig.~\ref{f:PScollage}, 
the $z-V_z$ plane is presented for 12 spherical volumes (4 kpc diameter) spread along the solar circle (Fig.~\ref{f:clock_xy2}). The volumes are numbered 1 to 12 in an anti-clockwise direction, which serves to identify the column in Fig.~\ref{f:PScollage}.
 After averaging over the phase-space volume, we form maps of $\langle V_\phi\rangle$ and $\langle V_R\rangle$. The numbered volumes are much larger than the original discovery volume (100 $\times$ 100 $\times$ 1000 pc).

Each row is a time step separated by 47.5 Myr intervals for the entire duration of the simulation (951 Myr), i.e. 21 time steps where the reference frame is the Rotational Standard of Rest (RSR). 
The simulated solar neighbourhood (volume 1) is in column 1.
Our focus on the solar circle is important for several reasons: (i) the local volume retains the clearest signal in the \gaia\ radial velocity data; (ii) to illustrate the phase spiral evolution, we remove the Galactic rotation that can only be carried out at one radius if we are to avoid complex resampling of the simulation.

The solar circle at the time of the
impact is shown in Region A (Fig.~\ref{f:PScollage}).
As already seen (Fig.~\ref{f:clock_xy}), the early stage evolution
reveals that the disc is highly disturbed up to at least $T=380$ Myr. At $T=238$ Myr, we observe a thick, red outer band (Region B) from a `feather' (in the language of L19), i.e. a part of the disc that separates and lifts away from the disc during the settling process in the outer disc.
A fixed amplitude, bending wave propagates along the filament; this leads to an outer ring or annulus in phase space as illustrated in Fig.~\ref{f:PSwaves}.

The phase spiral starts to emerge in both $\langle V_\phi\rangle$ and $\langle V_R\rangle$ simultaneously around $T=476$ Myr.
The phenomenon comes and goes at all radii from $R\gtrsim 3$ kpc to the outer disc. 
For the first time, we detect the phase spiral over the same extent in the $z-V_z$ plane and at the same intrinsic resolution as the discovery paper; earlier work (e.g. L19, B19) had intrinsically lower resolution and
a $z-$range twice as large compared to observations.
The inversion of the relative $z$ and $V_z$ extent as a function of radius is precisely as predicted in B19 (Fig. 20). But here the focus is on evolution along the solar circle, which is sufficient for our purposes.

In Fig.~\ref{f:PScollage}, diagonal bands from upper right to lower left are apparent in both mosaics. These track along the same line as the two sequences of yellow (left) and black (right) boxes.
The simulated solar neighbourhood (volume 1) lies along the same radius vector as the impact site at $T=95$ Myr. But this site increasingly lags behind volume 1 (period $T_\odot=210$ Myr) as the disc rotates, but realigns at a later time. The boxes indicate which volumes at specific timesteps are precisely aligned with the radius vector to the original impact site.

The period at the impactor radius ($R\approx 18$ kpc) is about $T_C=460$ Myr, such that volume 1 realigns with the impact site roughly every $\Delta T = T_{\odot,C}=(T_\odot T_C)/(T_C-T_\odot)\approx 380$ Myr. Interestingly, this is precisely the time delay for a phase spiral to emerge along the solar circle. The continued interaction of the solar neighbourhood with the region of the disc around
the impact site is a symptom of the wrapping up of the bending mode and the density wave. {\bf Thus, the spiral arms and bending mode are not strictly bisymmetric with respect to the Galactic Centre, implying that it may be possible to establish the transit time and its impact parameter in the Galactic disc with sufficient data, maybe even with the next \gaia\ RVS data release in 2022. The diagonal
banding in Fig.~\ref{f:PScollage} is a direct manifestation of this weak asymmetry.
}

In Fig.~\ref{f:PScollage}, we observe that the phase spiral first emerges at $T=476$ Myr.  In Region C at $T=523$ Myr, we see
the phase spiral clearly in the antipode region (cols. 7-9), something that is not seen diametrically opposite in the solar neighbourhood (cols. 1-3). The subsequent evolution tracks along the diagonal bands, and so is tied to the original impact site. Even though the density wave and bending mode appear bisymmetric by nature, the impulsive event is intrinsically one-sided. 

Fig.~\ref{f:bend-dens} shows how difficult it is to associate the phase spiral with the observed properties of the disc in the $x-y$ plane at any given epoch. The numbered $z-V_z$ diagrams around the outside are taken from row D (Fig.~\ref{f:PScollage}). The $x-y$ plane at $T=618$ Myr looks highly bisymmetric in projected stellar density, and in $\langle z\rangle$ and $\langle V_z\rangle$. But the phase spiral becomes stronger in both $\langle V_R\rangle$ and $\langle V_\phi\rangle$ as one moves around the clock (volumes 1 to 11).

As mentioned already, there is an inherent {\it asymmetry} in the strength of the disturbance, as we saw in Sec.~\ref{s:early}, in addition to the underlying $m=1$ mode. The phase spiral in each volume, to borrow from R.P. Feynman, is a sum over histories. The disc crossing occurred in the distant past but the transit region (albeit stretched by the disc's shear) lies immediately outside of volumes 8-9 at this time
(cf. Fig.~\ref{f:PScollage}). {\it The imprint of the one-sided disturbance is preserved for at least 500 Myr!}

At later times, with respect to the solar circle, the phase spiral develops both in the upstream (against rotation) and downstream direction (with rotation) due to the disc's shear. Once triggered, an increasing number of panels in Fig.~\ref{f:PScollage} show the phase spiral with the passage of time, and its fidelity also improves as more wraps emerge. By the
end of the simulation, the phase spiral has taken hold of the entire solar circle (Row E). In this respect, we can use the ubiquity of the phase spiral to age-date the impulsive event, and even to locate the earlier crossing point in the outer disc. These are themes that we anticipate will be addressed in future papers on Galactic seismology.

\begin{figure}
\includegraphics[width=0.5\textwidth]{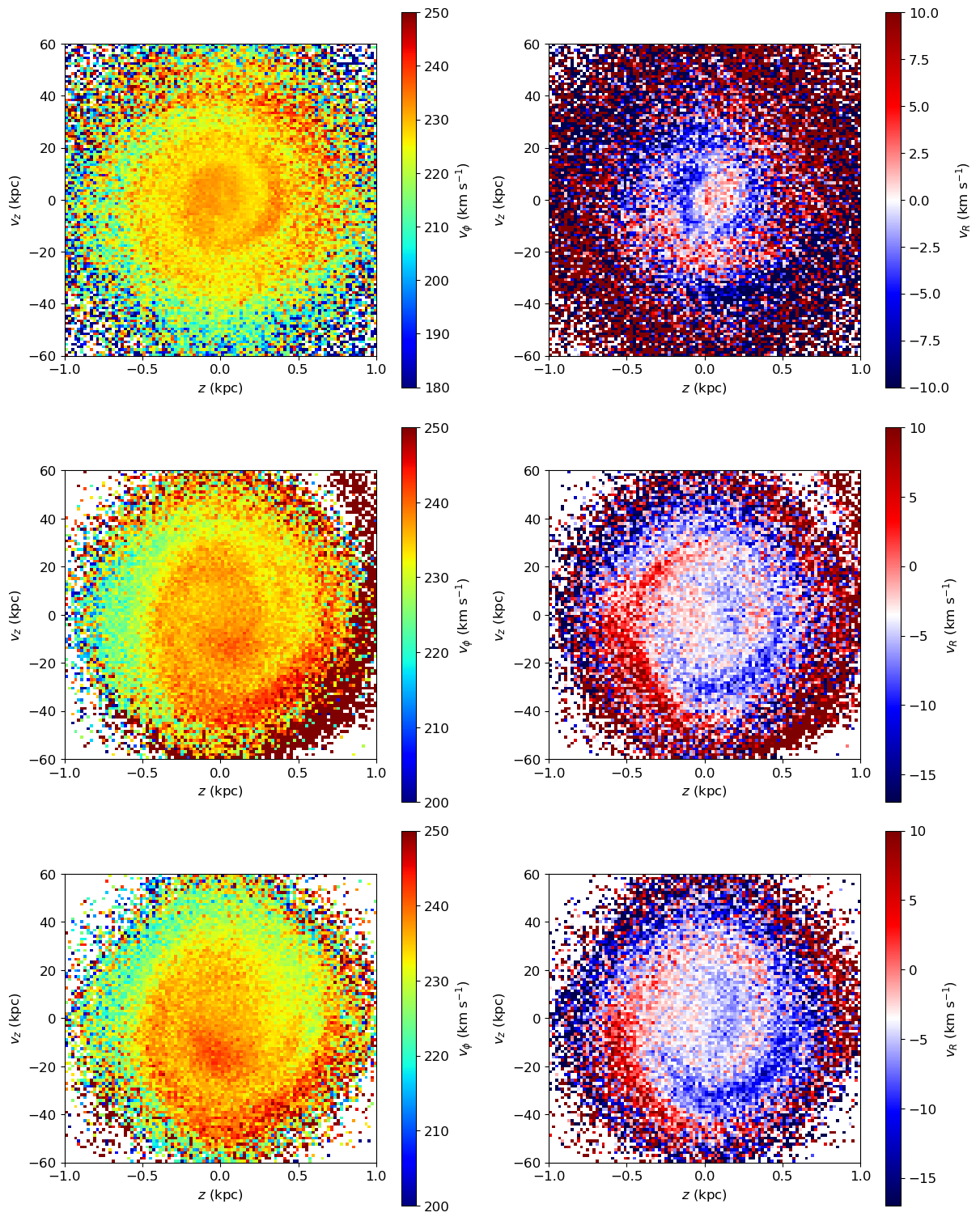} \caption{A comparison of the $z-V_z$ plane presented in the discovery paper \citep[][top]{antoja2018} with volumes 5 and 6 in time frame $T=951$ Myr (middle \& bottom); we have preserved the colour schemes from the earlier paper. Our volumes are much larger than the \gaia\ volume (see text) but we match the extent in the $z-V_z$ and the range in values for the first time; the fine structure of the phase spiral is evident in both also. The simulation produces fewer wraps in $\langle V_\phi\rangle$ (left) than in $\langle V_R\rangle$ (right), which may reflect the need for more disc particles and/or longer timespans in future simulations.
}
\label{f:antoja}
\end{figure}
\section{Discussion} \label{s:disc}

The \gaia\ discovery of the phase spiral was entirely unanticipated \citep{antoja2018}. As described in Sec.~\ref{s:model}, the most natural interpretation is incomplete phase mixing after a major disturbance of the disc. But the remarkable signature raises important questions: What is the precise mechanism that explains it? Is the signature fragile or robust? How widespread is it and can we learn from it?

On the balance of evidence presented, Galactic seismology, a subset of Galactic archaeology, has a rich future. But our work raises more questions than answers. In the closing comments of the last section, we make the case for follow-up studies.  Within the broad remit of Galactic seismology, an important goal is to age-date the origin of the phase spiral. Using two distinct mechanisms, B19 and \citet{Khoperskov2019} show that once the corrugation is excited, it can last up to about 1.5 Gyr. But repeated disc crossings from a massive perturber may wipe out the phase coherence imposed by the previous crossing (L19, B19). How are we to understand this?

{\bf
The orbit period today for the low-mass Sgr remnant is likely to be quite short \citep[$\sim$700 Myr,][]{Ibata1998}. In Sec.~\ref{s:intro}, we make the case for a high initial mass for Sgr $-$ of order the LMC mass ($\sim 10^{11}$ \Msun) $-$ for which, due to its orbit within the extended Galactic halo, dynamical friction is likely to be effective. The decay time and mass loss rate require knowledge of Sgr's initial orbital configuration \citep{Jiang2000}. But different orbit models lead to a catastrophic mass loss over a time frame of $7-9$ Gyr such that the observed low mass today is entirely plausible. In external galaxies, stellar streams have been modelled with mass loss rates of order 0.7 dex per orbit loop \citep{Amorisco2015}.
Thus, given the mass of Sgr today is too low to excite the phase spiral (Sec. 1), we suggest it was triggered by an earlier crossing in the past 1$-$2 Gyr and that Sgr has been losing mass at a high rate ($\sim 0.5-1$ dex per orbit loop) in the last few passages.
} 

Over timelines of order 1 Gyr, the phase-spiral signal appears to be quite fragile, coming into view in one time step (e.g. Fig.~\ref{f:PScollage}) and then either evolving {\it in situ}, migrating in azimuth (co-moving RSR frame), or fading altogether in the next time step. 
{\bf In our extended simulation run, we find that the last vestiges of the phase spiral are gone by $T=1.5$ Gyr.}
Although no panel in Fig.~\ref{f:PScollage} shows a close match in both $\langle V_\phi\rangle$ and $\langle V_R\rangle$ simultaneously, we find features that are ubiquitous and reminiscent of the discovery paper (see Fig.~\ref{f:antoja}). Note here that the simulated panels show weak evidence of a second phase spiral emerging out of sync with the dominant one. This can arise from having two phase-spiral arms moving through the same volume, which becomes increasingly likely as the density wave becomes highly wrapped at late times. 

There is a strong case for repeating the simulations with more particles (N $\gtrsim$ 10$^9$) allowing for smaller sample volumes. This would also help improve the simulation's declining spatial resolution at larger Galactic radius beyond $R \approx 10$ kpc (see Sec.~\ref{s:ics}).

The simulated phase spiral encoded by $\langle V_R\rangle$ is acceptable, but the phase spiral encoded by $\langle V_\phi\rangle$ shows up to one less wrap than observed. \cite{Khanna2019} found that the \gaia\ phase spiral is particularly prominent and well behaved in $\langle L_z\rangle$, but we were unable to see any improvement in our simulated signal when encoded in the same way. Given the central role of orbital angular momentum in generating the phase spiral (Sec.~\ref{s:BS18}), it is not clear why this should be. While our simulations achieve the recommended particle count of $N\sim 10^8$ \citep{Binney2018}, our test simulations with smaller intrinsic spatial resolution (Sec.~\ref{s:nbody}) did not alter the outcome significantly. But more particles would allow us to better match the \Gaia\ RVS volume and will be necessary for interpreting future \Gaia\ data releases.

There is one important consequence of our ``rollercoaster'' model that has been glossed over. The interaction of density waves and bending modes implies that the younger stellar populations are those that are more likely to be caught up in the phase-spiral action. It is probable that older stars also contribute to the signal, but the feature is likely to be more enhanced for the younger stars. While there is existing evidence that this is true \citep[B19,][]{Li2020}, the dominant population of the phase spiral needs to be addressed more carefully (e.g. better stellar ages) in future studies \citep{Sharma2016,Miglio2020}.

This brings us full circle to the original prediction of the phase spiral (with two arms rather than the singular arm observed by \gaia) arising from dissolving star clusters \citep{Candlish2014} that predated the \gaia\ era. Spiral arms in the Milky Way can be identified from star-forming regions \citep[e.g.][]{Lepine2001}, as they can be in external galaxies, but this is no guarantee that young star clusters ($\lesssim 100$ Myr) are entirely restricted to the spiral arms. An inventory of almost 2000 star clusters confirmed by \gaia\ \citep[][Fig. 8]{CantatGaudin2020}
shows that star clusters of all ages inhabit our neighbourhood within a few kiloparsecs, with the youngest star clusters confined to broad bands rather than ridge lines. Most clusters that form appear to dissolve within 100 Myr \citep{Krumholz2019,Adamo2020} such that cluster dissolution can conceivably contribute to the innermost wraps of the phase spiral \citep[cf.][]{Li2020}. Cluster dissolution is unlikely account for the signature {\it in toto}, not least because the models to date predict a 2-arm phase spiral \citep{Candlish2014}, and that has not been observed so far.

Intriguingly, \cite{Michtchenko2019} identify parts of the local phase spiral as arising from several moving groups, including Pleaides, Hyades, Sirius (Ursa Major) and Coma Berenices. These are four of the closest clusters, all of which appear to be unbound. What is particularly striking is that their ages range from 100 to 600 Myr \citep{Famaey2008} with the youth carrying the stronger signal.
More work is needed to assess the detailed structure of the local corrugation \citep[e.g.][]{Friske2019,Beane2019} and its relationship to the local spiral arms \citep[e.g.][]{Miyachi2019,Griv2020}. How these relate to the local young clusters and star-forming regions also becomes important \citep[][]{Quillen2020,CantatGaudin2020}.

It is likely that our cold-disc simulations, while very carefully constructed, are too simplistic.
{\bf \cite{GrionFilho2020} have called for a holistic approach to modelling and understanding galaxy interactions. In this respect, there is a great deal to be done.
\cite{Vasiliev2021} have indicated that the LMC's interaction with the Galaxy and Sgr are important to address, particularly in relation to the development of Sgr's tidal loops.
In the next phase, simulations must introduce a clumpy dark matter halo, giant molecular clouds and smooth gas, a central bar and internally-driven spiral arms, with the overarching goal of producing a more realistic framework for future study.} This work will inform the nature of future studies, leading to better targetted surveys rather than all-sky surveys to test different non-equilibrium dynamical models. 

To date, we have not considered the contribution of the cool gas. 
Another prediction from B19 (their Sec. 8.2) is that the corrugation may exist in the Galactic \HI\ or the molecular hydrogen distribution, and may even explain Gould's belt and other local phenomena.
{\bf Intriguingly, such a gas wave may already be evident \citep{Alves2020}. Connecting this information with star clusters at different stages of their evolution may also provide important insights \citep[e.g.][]{Quillen2020}. If the spiral arms and bending modes are triggered by Sgr, we predict that the corrugations of the wrapping-up wave pattern do {\it not} align with the spiral arm pattern
(e.g. Fig.~\ref{f:wrap}).
}

In the spirit of \cite{Iye1985} and \cite{Widrow2012}, there is a case for extending galactoseismology to the study of nearby galaxies \citep[cf.][]{Gomez2020}. Arguments have been made for undulations in the Galactic rotation curve \citep{MartinezMedina2019} being correlated with the spectacular $R-V_\phi$ ridges seen by \gaia\ \citep{antoja2018,Khanna2019}, or possibly even with the spiral arms \citep{Williams2013}. With careful disc modelling and subtraction, it may be possible to identify the signatures of bending modes in substantial numbers of nearby disc galaxies \citep[cf.][]{Matthews2008b,Gomez2020}. In such instances, it will be important to establish whether the bending mode wraps up with the disc rotation, as we show here. Morphologically, the wrapping of the spiral density wave is more easily determined and thus it may even be possible to establish whether the interaction of these two wave modes is a regular phenomenon.

{\bf
\section{Summary of main results}
\label{s:summary}

The main goal of this work is to provide a better understanding of the response of the Galactic stellar disc to a strong impulse in an N-body setting. Our work shadows \cite{Binney2018} who use an analytic model of a disc-crossing satellite ($2\times 10^{10}$ \Msun) to explain the phase spiral and its coupled behaviour in $\langle V_R\rangle$ and $\langle V_\phi\rangle$. 
Here are the main findings:

\begin{enumerate}
    \item The initialisation code \agama\ \citep{Vasiliev2018,Vasiliev2019}
    is essential to setting up long-term stability in isolated, multi-component, fully self-consistent
    realizations (defined at the outset) of the Galaxy, as we demonstrate here. We consider this to be a major step in the field of Galactic seismology. 
    
    \item As is well known since the advent of N-body codes, a disc-crossing mass drives spiral arms that wrap up as the disc rotates differentially. More recent work has revealed the vertical, wave-like response of the cold disc in addition to the spiral arm perturbation. Here we find that the impulse triggers a superposition of two distinct bisymmetric ($m=2$) modes $-$ a density wave and a bending wave $-$ that wrap up at different rates. Stars in the {\it faster} density wave wrap up with time $T$ according to $\phi_D(R,T)=(\Omega_D(R) + \Omega_{\rm o})\:T$ where $\phi_D$ describes the spiral pattern and $\Omega_D =\Omega(R) -\kappa(R)/2$. While the pattern speed $\Omega_{\rm o}$ is small, it is non-zero indicating that this is a dynamic density wave. The {\it slower} bending wave wraps up according to $\Omega_B\approx\Omega_D/2$ producing a corrugated wave, a phenomenon now well established in the Galaxy. 
    
\item The bunching effect of the density wave triggers the ``phase spiral" in the $z-V_z$ plane as it rolls up and down on the bending wave (``rollercoaster'' model). In other words, the wrapping-up spiral arms undulate with the slower wrap of the bending mode, producing a rollercoaster effect {\it along each spiral arm}. This non-equilibrium system is capable of driving the phase-spiral phenomenon.

    \item In agreement with earlier work \citep[e.g.][]{antoja2018}, the phase spiral is most evident when each point of the $z-V_z$ phase plane is represented by either $\langle V_R\rangle$ or $\langle V_\phi\rangle$, averaged over the volume, indicating that the disturbance has locked the vertical oscillations with the in-plane epicyclic motions. We find that the phase-spiral action takes about $\Delta T\approx 400$ Myr to become a disc-wide phenomenon, but once it sets in, it can survive till about 1.5 Gyr after the disc transit. (The spiral arms and the bending mode are not strictly bisymmetric with respect to the Galactic Centre: the time delay $\Delta T$ reflects the time it takes the solar neighbourhood and the impact site in the outer disc to realign radially.)
    
    \item In earlier work (B19, L19), it was shown that a massive perturber ``resets'' the disc by wiping out any existing phase-spiral dynamics from an earlier disc transit. Thus, given the mass of Sgr today is too low to excite the phase spiral (Sec.~\ref{s:intro}), we suggest it was triggered by an earlier crossing in the past 1$-$2 Gyr and that Sgr has been losing mass at a high rate ($\sim 0.5-1$ dex per orbit loop) in the last few passages.
    
\end{enumerate}

}

\section*{Acknowledgements}

JBH is supported by an ARC Australian Laureate Fellowship (FL140100278) and the ARC Centre of Excellence for All Sky Astrophysics in 3 Dimensions (ASTRO-3D) through project number CE170100013. JBH \& TTG acknowledge the resources and services from the National Computational Infrastructure (NCI), which is supported by the Australian Government. We thank E.~Vasiliev for his valuable assistance and continued conversation with the use and development of the \agama\ code.  We thank
James Binney for earlier conversations that forced
us to look at \agama\ carefully. We are indebted to Romain Teyssier for assisting us with the idea of a time-dependent impulsive mass, both in terms of our additions to the code, and for allowing sink particles for the first time to operate with pure N-body simulations. We thank Shourya Khanna and Sanjib Sharma for providing the \Gaia\ DR2 data necessary to recreate the discovery images from \cite{antoja2018}.
We are particularly grateful to Ken Freeman for many years of collegiality and for insightful and helpful comments on the manuscript. This work has made used of Matplotlib, a Python-based plotting package \citep{hun07a}, Pynbody \citep{pynbody} and Gnuplot, originally written by Thomas Williams and Colin Kelley.

\section*{Data availability}

The \ramses\ and \agama\ codes were developed by independent investigators and are readily available online. Our work concentrated on designing suitable setup files to run with these facility codes; these are freely available upon request. All simulated data sets specific to the figures are also available upon request.


\input{phasewrap.bbl}

\appendix

{\bf
\section{Brief history of Galactic seismology}\label{s:history}

Galactic seismology explores the response of the Galaxy to internal and external perturbations. This is where we learn about Galactic structure, the role of self-gravity and dark matter within the Milky Way, inter alia. The specific language of `galactic seismology' was first used by \cite{Iye1985}. In later years, \cite{Widrow2012} coined the term `galactoseismology' to describe their related analysis of N-body simulations of galaxy discs; we consider both terminologies to be synonymous. Here we clarify how the field has emerged; it can be considered as a subdiscipline of the more expansive field of Galactic archaeology.

It has been known for a century \citep{Jeans1915} that there are common elements between equations that govern
the dynamics of continuous fluids and those that describe discrete stellar systems, even while they obey different equations of state \citep[][]{Binney2008}: To paraphrase, the analogies arise because a fluid system is supported against gravity by gradients in the isotropic scalar pressure ($-\nabla p$), while a stellar system is supported by gradients in the stress tensor ($-\nu\sigma^2_{ij}$; $\nu$ is the local density and $\sigma$ is the local velocity dispersion for spatial coordinates $i,j$) that describe anisotropic pressure. The dynamical response of a discrete or a continuous medium to internal or external perturbations can be described in terms of propagating wave modes (wavelength $\lambda$). These in turn can be examined for local stability using dispersion relations to assess the relative roles of growing and decaying modes of a given wavenumber ($k=2\pi/\lambda$).

The study of global modes as applied to a linear analysis of disc galaxies goes back half a century \citep{Hunter1963,Hunter1965,Hohl1971,Kalnajs1972,Bardeen1975,Iye1978,Takahara1978}.
This work intensified in the critical response to the \cite{Lin1966} theory for describing galactic spiral structure \citep[e.g.][]{Hunter1969,Binney2008}.
Notably, Hunter introduced methods based on Legendre polynomial 
expansions to study the global oscillations of a cold, self-gravitating disc.
By studying the low-order spatial eigenmodes across a finite disc, the task
is reduced an eigenvalue problem \citep[cf.][]{Yabushita1969}.

In his short paper, \cite{Iye1985} formulates the language of `galactic seismology' for the first time with a
view to the emerging field of asteroseismology. As a historical note, this was the same year asteroseismology was formally declared to be a subdiscipline of astrophysics \citep{Aerts2021}.
Iye states that ``it is natural to 
expect that the oscillations of a rotating gas disc (e.g. galaxy) share a good
deal of physical similarity with those of a rotating gas sphere (e.g. star).'' He even describes the dominant modes of a rotating gas disc in a language familiar to all stellar seismologists. With reference to Hunter's work 
\citep{Aoki1978,Aoki1979},
Iye and colleagues apply modal analysis techniques to galaxy observations for the first time \citep{Iye1983,Ueda1985},
although now specific references to $p$ (pressure) modes and $g$ (gravity) modes are no longer evident.

\cite{Widrow2012} introduced the language of `galactoseismology,' bolstering the etymological association with asteroseismology \citep{Gough1996}. This terminology was introduced in recognition of wave-like patterns in the Galactic stellar disc
observed in both star counts and stellar kinematics by this group \citep[see  also][]{Williams2013}.
They did not include modal analysis in their brief study, other than a statement of its importance to the problem, although a more sophisticated study was soon to follow
\citep{Widrow2014}. More recent papers by Widrow and collaborators explore these
ideas in more detail, specifically with reference to N-body simulations
\citep[q.v.][]{Chequers2018}.
}

\section{Impulse approximation}\label{s:impulse}

\begin{figure}
\includegraphics[width=0.5\textwidth]{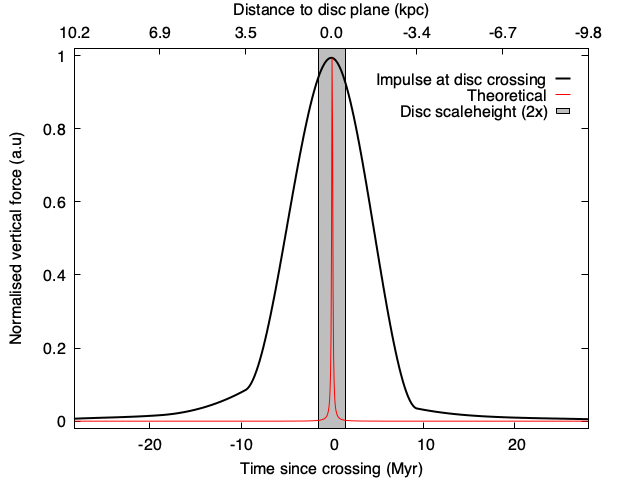} \caption{The force experienced by a star close to Sgr's crossing point ($R \approx 18$ kpc) as a function of time (solid curve).  This is taken from the N-body simulation data dumped to storage roughly every 10 Myr. Since the time step in the simulation is orders of magnitude smaller, the impulsive response to the intruder is vastly narrower in the simulation. This is illustrated by the red curve, which indicates the theoretical result. The small asymmetry imposed by time-dependent intruder mass is clear. Note that the curves have been normalised to a peak height of unity. The shaded area indicates the width of the stellar disc defined as twice its scale height ($2z_t \approx 500$ pc).}
\label{f:impulse}
\end{figure}

Our set-up ensures that we are working entirely within the impulse approximation. In Fig.~\ref{f:impulse}, we show the force experienced by a star in the immediate vicinity of Sgr's crossing point ($R \approx 18$ kpc) as a function of time before and after Sgr's disc transit (solid curve). The top $x$-axis displays Sgr's vertical distance to the disc plane along its orbit. The gray-shaded area indicates the width of the stellar disc, here given by twice its scale height ($2z_t \approx 500$ pc). The interaction between stars at the crossing point is clearly impulsive. The apparent asymmetry in the impulse is the result of the reduced mass after transit (cf. Sec.~\ref{s:model}). The impulsive force in the simulation is much narrower in practice, but appears broadened due to the coarse time resolution ($\Delta t \approx 10$ Myr) adopted for storing the timesteps. To emphasize this point, we include in the figure the theoretical result (red curve). The latter corresponds to the force experienced by a body in the vicinity of a point mass, softened by $\epsilon_s$, i.e. $\sim r / (r^2 + \epsilon_s^2)^{3/2}$ where $r$ is the radial separation.

\label{lastpage}
\end{document}